\documentclass[a4paper,11pt]{article}
\pdfoutput=1 
\usepackage{jcappub} 
\usepackage[T1]{fontenc} 

\usepackage{amssymb}
\usepackage{amsopn}
\usepackage{hyperref}
\usepackage{xcolor}
\usepackage{mathtools}
\usepackage[caption=false]{subfig}
\usepackage{soul}
\usepackage{amsmath}
\usepackage{graphicx}
\usepackage{float}
\usepackage{cleveref}
\usepackage[parfill]{parskip}
\usepackage{amsopn}
\usepackage{braket}
\usepackage{slashed}
\usepackage{bbold}

\usepackage{mathtools,leftidx} 

\title{\boldmath Cosmology with explicit and spontaneous background fields}
\author[a,1]{Carlos M. Reyes,  }
\author[a,b]{C\'esar Riquelme}
\author[c,2]{and Alex Soto} 
\affiliation[a]{  Centro de Ciencias Exactas, Universidad del B\'{i}o-B\'{i}o,
 Chill\'{a}n, Casilla 447, Chile}
\affiliation[b]{Departamento de F\' {\i}sica, Universidad de Concepci\'on, Casilla 160-C, Concepci\'on, Chile}
\affiliation[c]{School of Mathematics, Statistics and Physics, Newcastle University, Newcastle upon Tyne, NE1 7RU, UK}
\emailAdd{creyes@ubiobio.cl}
\emailAdd{ceriquelme@udec.cl}
\emailAdd{alex.soto@ncl.ac.uk}
\abstract{We study a general class of effective
backgrounds that break diffeomorphism invariance and their potential roles in cosmology.
Specifically, we examine both explicit and spontaneous background fields which display distinct 
transformation properties and are 
characterized with different dynamics.
For explicit breaking, we focus on the $t$-sector of the
minimal gravitational Standard-Model Extension (SME) and for spontaneous breaking on a
vector field called the  bumblebee model.
In both cases, we derive the modified Friedmann
equations and find a configuration of the background fields that preserve 
isotropy and homogeneity.
We show that the explicit $t$-sector admits phases of accelerated expansion of the universe 
with standard matter.}
\begin{document}
\maketitle
\flushbottom
\section{Introduction}
The $\Lambda$CDM model, or concordance cosmology has been
constructed over a wide-ranging and extensive set 
 of observational data. 
The model consists in a hot big bang
 with a tiny positive cosmological constant, cold dark matter
and an initial period of rapid expansion known as inflation. 
The $\Lambda$CDM model fits extremely well with many observations, in particular with  
 the
Hubble law of recession of galaxies, the cosmic microwave 
background (CMB), 
and the abundance of light elements according to
primordial
nucleosynthesis~\cite{Peebles:1966rol,Walker:1991ap,Coc:2006sx}.
Considering the observed galaxy distribution~\cite{SDSS:2005xqv} and the measurements of CMB
anisotropy~\cite{WMAP:2003zzr,WMAP:2003elm}, the assumptions of spatially isotropy and homogeneity 
are good approximations for our universe.
Paradoxically, although dark matter constitutes the largest contribution of matter in the universe,
it has not been directly detected~\cite{Smith:1988kw}. Nevertheless, it
plays a crucial role in the 
formation of large-scale structure of galaxies~\cite{SDSS:2005xqv,Daniel:2008et,BOSS:2016wmc}, temperature fluctuations 
and polarization patterns in the CMB~\cite{WMAP:2008lyn,WMAP:2010qai,WMAP:2012nax,Planck:2015fie,Planck:2018vyg}. 

Despite its observational success, the $\Lambda$CDM model is considered an effective theory
 requiring extensions at both the smallest distances scales in pre-inflationary times and at the largest distance scales. During the early 
 universe, when it was infinitely dense and hot, quantum fluctuations of the gravitational field 
 become significant, necessitating a quantum gravity theory which is still a challenge. On the largest scales, the standard 
 cosmological model poses numerous questions. 
 One question concerns the small value of the cosmological constant, responsible for the current acceleration of the universe~\cite{SupernovaSearchTeam:1998fmf,SupernovaCosmologyProject:1998vns,SupernovaCosmologyProject:2008ojh,Pan-STARRS1:2017jku}, remaining an unresolved issue, often attributed to an unknown form of vacuum energy or dark energy~\cite{Turner:1998mg}.
To address some of these theoretical challenges, several modifications of gravity and effective
 theories have been proposed. In particular, dark matter candidates have been suggested in the
  form of WIMPs~\cite{Hut:1977zn,Lee:1977ua,Ahlen:1987mn}, including axions~\cite{Ellis:1998gt} 
  and particles predicted by supersymmetry~\cite{Ellis:1983ew}. For dark energy, various models 
  have been proposed, such as quintessence~\cite{Caldwell:1997ii,Kolb:1981hk}, k-essence~\cite{Armendariz-Picon:1999hyi,Garriga:1999vw,Armendariz-Picon:2000nqq,Armendariz-Picon:2000ulo}, Galileons~\cite{Nicolis:2008in,Pirtskhalava:2015nla,deRham:2021fpu,Nascimento:2020jwx}, 
  massive gravity~\cite{deRham:2010ik,deRham:2010tw}, chameleons~\cite{Khoury:2003rn}, tensor-vector models~\cite{Ford:1989me,Armendariz-Picon:2004say,Wei:2006tn,BeltranJimenez:2008iye}, and fluid matter~\cite{Fabris:2001tm,Cruz:2006ck}. 
  For a comprehensive overview, see the review~\cite{Sapone:2010iz}.

Alternative approaches to address these ultraviolet and infrared issues have been 
proposed in the form of effective theories that incorporate the breaking of diffeomorphism 
and local Lorentz symmetries. Generally, these theories include tensorial background fields 
that emerge from an underlying theory, presumably valid at Planck scales. Notably, the mechanism 
of spontaneous diffeomorphism symmetry breaking has been shown to arise from string 
theory~\cite{Kostelecky:1988zi,Kostelecky:1991ak,Kostelecky:1995qk}. This possibility has 
led to the development of the generalized effective framework known as the Standard-Model Extension (SME).
The SME accommodates arbitrary dimension operators that describe small departures from 
local Lorentz and diffeomorphism symmetries, incorporating both nondynamical and dynamical 
background fields. The framework distinguishes between a minimal sector, which includes only 
renormalizable operators, and a nonminimal sector, which includes operators of mass dimension 
higher than four. Numerous bounds on local Lorentz and diffeomorphism violations have been 
established for both matter and gravity~\cite{Kostelecky:2008ts}.

The gravitational sector of the SME were presented in \cite{Kostelecky:2003fs}, in the framework 
of Riemann-Cartan spacetime, including as limiting cases the Riemann and Minkowski geometries.
 In general, one identifies two classes of local Lorentz and diffeomorphism symmetry violations
 which are produced by explicit and spontaneous background fields.
In both cases
  the background fields are responsible for the
 breaking of local Lorentz and diffeomorphism invariance as they 
 transform covariantly under 
general coordinate transformations but remain fixed under particle transformations~\cite{CK1,CK2,Bluhm:2014oua}.
The breaking of explicit diffeomorphism is
 performed with nondynamical backgrounds terms included in the effective Lagrangians, which usually are coupled non-minimally
 to gravity. These backgrounds 
 have been used to study
the post-newtonian limit~\cite{Bailey:2006fd}, gravitational torsion~\cite{Kostelecky:2007kx}, 
 static and spherically symmetric solutions~\cite{Bonder:2020fpn}.
Spontaneous backgrounds came from an underlying theory
in which the fields acquire a finite vacuum expectation through the potential.
An emblematic spontaneous background is the bumblebee field that has been studied 
for radiative corrections~\cite{Delhom:2022xfo,Delhom:2020gfv,Nascimento:2023auz}, cosmology~\cite{Armendariz-Picon:2009kfd,Jesus:2019nwi},
G\"odel-type universes~\cite{Jesus:2020lsv,Santos:2014nxm}, asymptotic flatness~\cite{Bonder:2021gjo}, the Kalb-Ramond field~\cite{Altschul:2009ae}, Einstein-aether models~\cite{Jacobson:2000xp,Armendariz-Picon:2010nss}.
Some consequences of spontaneous diffeomorphism violation are the arising of 
 Nambu-Goldstone and Higgs modes~\cite{Bluhm:2004ep,Bluhm:2007bd}.
Background fields have been also used in 
other contexts in gravity, see~\cite{deRham:2010kj,Katz-BT,Deruelle:2017xel,Armendariz-Picon:2023gyl,Gamboa:2022msv}. 

Explicit diffeomorphism violation has been showed to be more difficult to implement 
due to possible conflicts between dynamics and geometry; which also has been called the no-go result which follows from the 
Bianchi identity~\cite{Kostelecky:2003fs,Bluhm:2014oua,Bluhm:2016dzm,Kostelecky:2020hbb,Bluhm:2021lzf,Bluhm:2023kph}.
Some elaborations have been proposed to deal with this issue such as the Stuckelberg method~\cite{Bluhm:2016dzm,Bluhm:2019ato,Bluhm:2023kph} or extending to 
 Finsler geometries~\cite{Kostelecky:2021tdf,Schreck:2014hga,Schreck:2015seb}. 
The condition involving the Bianchi identity,
the non invariance of particle transformation of the action
and possible geometry constraints has been a challenge to solve.
Interesting, this condition can be surmounted
in the Chern-Simons modification of gravity by restricting the geometry with the condition
$ \prescript{*}{ }{R}R =0 $, hence, restoring the diffeomorphism invariance dynamically~\cite{Jackiw:2003pm}.
In some sectors of the minimal gravitational SME, recently it has been shown a similar condition involving vacuum solutions $R=0$
for strong gravitational systems~\cite{Bailey:2024zgr}.
In our work, we explore alternative solutions to the consistency condition 
in the context of cosmology where a maximally symmetric 
subspace arises.

Recently, the minimal sector of the gravitational SME has been casted into canonical form~\cite{ONeal-Ault:2020ebv,Reyes:2021cpx}
and shown to produce consistent Hamilton-Jacobi equation of motion~\cite{Reyes:2022mvm}.
Models with nondynamical background fields involving the
 $u$ and $s^{\mu \nu}$ fields have been studied for Friedmann spacetimes in cosmology including searches 
 for cosmic speedup~\cite{Reyes:2022dil,Nilsson:2023exc}, primordial fluctuations~\cite{Nilsson:2022mzq}, 
 the Hubble tension~\cite{Khodadi:2023ezj}, and boundary properties~\cite{Reyes:2023sgk}.
 We continue these studies and 
 generalize to cover the $t$-sector and also to include the bumblebee dynamical field
 in our approach.

The article is organized as follows, in section~\ref{SectionII} we introduce the $3+1$
formalism, discuss the properties of background fields and develop a technique to 
consistently implement the diffeomorphism violation 
in the context of cosmology.
In section~\ref{SectionIII} we consider explicit diffeomorphism breakdown. In particular, we 
present the gravitational SME model, focus on the $t$-sector and derive the
 boundary terms, derive the modified FRWL equations and study possible mechanisms for 
 late acceleration in the universe. In 
section~\ref{SectionVI} we consider spontaneous diffeomorphism violation 
using the bumblebee model, finding the modified FRWL equation and focusing on some specific sectors
of the background. To organize properly the work we have considered four appendices with the details of the calculations, these are appendices
\ref{AppendixA}, \ref{AppendixB}, \ref{AppendixC}, \ref{AppendixD}.
The last section~\ref{SectionVIII} contains our final remarks.
\section{Basics}\label{SectionII}
In this section, we provide a brief introduction to the 
$3+1$ formalism, where spacetime is decomposed into space and time~\cite{Arnowitt:1962hi}.
A slight depart is taken by working 
with spatial evaluated tensors $e_a^{\mu}$ and $\widetilde{E}^a_{\mu}$, defined in 
 the hypersurface $\Sigma_t$. This allows to decompose the modified Einstein's equation
 and to find the Friedmann equations in a simplified manner as given in the next sections.
We also discuss explicit and spontaneous diffeomorphism symmetry breaking 
and the potential
issues that appear in the case of explicit breaking. 
The final part of this section focuses on identifying a subset of diffeomorphism-breaking coefficients
that preserve the symmetries of isotropy and homogeneity.
\subsection{The $3+1$ formalism}\label{SectionI:SubsectionA}
We start by introducing $4$-dimensional spacetime manifold $\mathcal M$ endowed 
with metric tensor $g_{\mu \nu}(x)$
and local coordinates $x^{\mu}$. We introduce a scalar function of coordinates 
$t(x^\mu)$, and  
foliate  $\mathcal M$ 
 into a family of hypersurfaces $\Sigma_t$ at constant time $t$. We require 
 the hypersurfaces to be non-intersecting and spacelike. Hence, the normal vector field 
 on $\Sigma_t$
 points towards the future, satisfying  $n^{\mu} \partial_{\mu} t > 0 $, which can be written as
 \begin{align}
n_\mu = -N \partial_{\mu} t \,,
 \end{align}
where the normalization factor
$N=\left |g^{\mu \nu } \partial_{\mu } t \partial_{\nu  }t  \right |^{-1/2}$ is called the lapse function.
We also can show that
\begin{align}
n^\mu n_\mu = -1\,.
 \end{align}
\begin{figure}[h]
\centering
\includegraphics[scale=0.5]{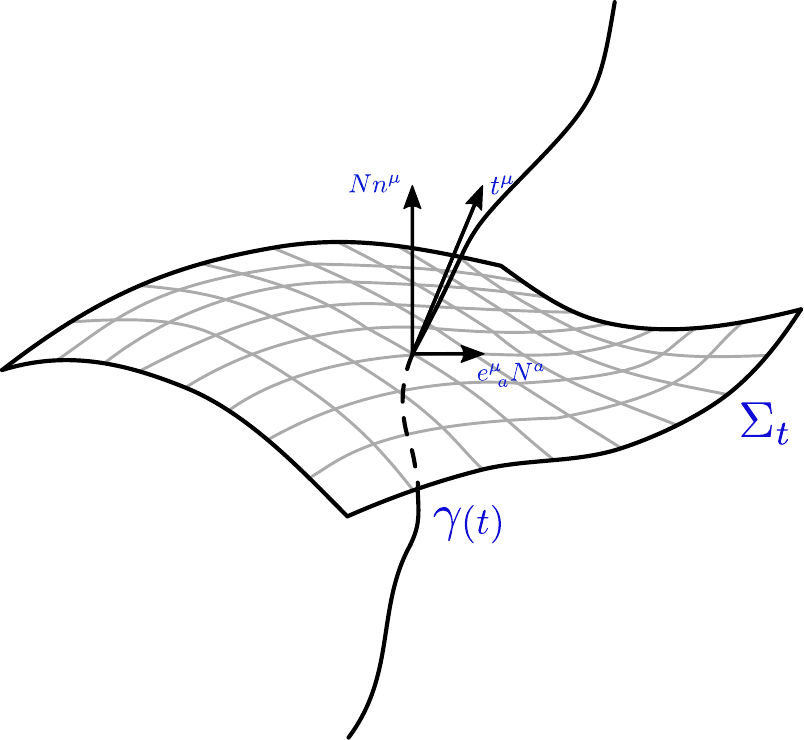}
\caption{Decomposition of spacetime $\mathcal M$ into space and $\Sigma_t$ time $t$, based on the congruence of 
curves $\gamma(t)$ and in terms of normal $n^{\mu}$ 
and tangent vector fields $e^{\mu}_a$.}
\label{Fig1}
\end{figure}
We continue with installing a set of local coordinates in space $y^a, y'^a, \dots ,y^{(n)a}$
on each hypersurfaces of the foliation $\Sigma_t, \Sigma'_t,  \dots, \Sigma^{(n)}_t $, respectively, where the indices are
 $a=1,2,3$. In order to relate each hypersurface,
 we introduce a congruence of curves $\gamma(t)$ that pinches the hypersurfaces at only one spatial 
 point, namely ${\bar y}^a={\bar y}'^a=\dots,={\bar y}^{(n) a} $.
This construction induces a transformation from the original coordinates $x^{\mu}$ on $\mathcal M$ to a one defined by
time $t$ and space $y^a$ given in the form
 $x^{\mu}=x^{\mu}(t,y^a)$, see the Ref.~\cite{Poisson:2004}.

We define the flow of time represented by the tangent vector to $\gamma(t)$ 
\begin{subequations}
\begin{align}
t^{\mu} &:=\left(\frac{\partial x^{\mu}}{\partial t}\right)_{y^a}\,,
\end{align}
and the vector in the tangent space of $\Sigma_t$
\begin{align}
   e^\mu_{a}&:=\left( \frac{\partial x^{\mu}}{\partial y^a} \right)_t\,.
\end{align}
\end{subequations}
The label in the parenthesis 
means at constant $y^a$ and $t$, respectively, see Fig.~\ref{Fig1}.

Considering displacements in the three-manifold, we define the induced metric $q_{ab}$ or first fundamental form 
 on $\Sigma_t$ by
\begin{align}
   q_{ab}:=g_{\mu\nu} e^\mu_{ a}e^\nu_{ b} \,.
\end{align}
and considering variations of the normal, follows 
 the extrinsic curvature or the second fundamental form
\begin{align}\label{Extrinsic}
K_{ab}=   e^{\mu} _a e^{\mu} _b  K_{\mu \nu} \,.
 \end{align}

To find the decomposed spacetime metric we consider
\begin{align}
dx^{\mu}= \left(\frac{\partial x^{\mu}}{\partial t}\right)_{y^a} dt+\left( \frac{\partial x^{\mu}}{\partial y^a} \right)_t  dy^a     \,,
\end{align}
and write $t^{\mu}$ 
in terms of normal and tangential directions as
\begin{align}\label{tangent-decomp}
t^{\mu}=Nn^{\mu}+N^a e^{\mu}_{a} \,, 
 \end{align}
where $N^a$ is 
called the shift vector, and replace above to obtain
\begin{align}
dx^{\mu}= n^{\mu} (N dt) +e^{\mu}_{a} (N^a dt+ dy^a)\,.
\end{align}
From the line element $ds^2=g_{\mu \nu} dx^{\mu}  dx^{\nu} $, 
we identify the metric components
 \begin{subequations}
\begin{align}
 g_{00}= -N^2+q_{ab}N^a N^b\,, \quad g_{0a}= N_a\,,  \quad  g_{ab}= q_{ab}\,,
  \end{align}
 and its inverse components
 \begin{align}
  g^{00}= -\frac{1}{N^2}\,, \quad  g^{0a}= \frac{N^a}{N^2}\,, \quad  g^{ab}= q^{ab}-\frac{N^a N^b}{N^2}\,.
 \end{align}
  \end{subequations}
Also, we can show that  $\sqrt{-g}=N\sqrt {q}$, which relates spacetime and induced 
metric determinants.

The property of orthogonality $e^\mu_{a}n_\mu=0$,
leads to the completeness relation
for the inverse metric
\begin{align}
    g^{\alpha\beta}=q^{ab}e^\alpha_{ a}e^\beta_{b}-n^\alpha n^\beta,
\end{align}
and to the expressions
 \begin{subequations}
\begin{align}\label{delta_strat}
    \delta^\alpha_{ \beta}&=e^\alpha_{ a}(g_{\beta\gamma}e^\gamma_{ b}q^{ba})-n^\alpha n_\beta\,,  \\
    g_{\alpha\beta}&=(g_{\alpha\rho}e^\rho_{ a}q^{ac})q_{cd}(q^{db}
    e^\sigma_{ b}g_{\sigma\beta})-n_\alpha n_\beta \,.
\end{align}
 \end{subequations}
This motivates the definition of the object
\begin{align}
    \widetilde E_\alpha^{ a}:=g_{\alpha\beta}q^{ab}e^\beta_{ b}\,,
\end{align}
which satisfies the following properties 
 \begin{subequations}
\begin{align}
    n^\alpha \widetilde E_\alpha^{ a}&=0\,, \\
    e^\alpha_{ b} \widetilde E_\alpha^{ a}&=\delta^a_{\ b} \,,
    \\
    e^\alpha_{ c}\widetilde E_\beta^{ c}&=\delta^\alpha_{ \beta}+n^\alpha n_\beta\,.
\end{align}
 \end{subequations}
 The new tensor $\widetilde E_\mu^{ a}$ acts as the inverse of $e^{\alpha}_b$ in the contracted spacetime
  indices but as a projector $q^{\alpha}_{\beta}$ in contracted spatial indices. Recall the projector 
  has properties $q^{\alpha }_{\beta} q^{\gamma }_{\alpha}=q^{\gamma }_{\beta}$ with $q^{\alpha }_{\beta}=\delta ^{\alpha}_{\beta}+  n^{\alpha} n_{\beta}$.  
  
In this way the completeness relations can be written as
 \begin{subequations}
\begin{align}
    g^{\alpha\beta}&=e^\alpha_{a}e^\beta_{b}q^{ab}-n^\alpha n^\beta \,, \label{comp_invmetric} \\
\delta^\alpha_{ \beta}&=e^\alpha_{a} \widetilde E_\beta^{ a}-n^\alpha n_\beta \,,   \label{comp_delta}   \\
g_{\alpha\beta}&=\widetilde E_\alpha^{a}   \widetilde E_\beta^{ b} q_{ab} -n_\alpha n_\beta    \label{comp_metric}  \,.
\end{align}
 \end{subequations}
We present 
 several expressions involving the new object $\widetilde E^a_{\mu}$ in the Appendix~\ref{AppendixA}.
\subsection{Symmetries and background fields}\label{SeccionII:Subsection:C}
The role of diffeomorphism invariance is fundamental in the construction of GR. 
On the one hand, it ensures the covariance of the 
 equations of motion and on the other, and more profoundly, it
reveals the geometric character of the theory.
This becomes specially apparent in the canonical formulation when 
using constraint generators of spacetime symmetries to 
 smoothly 
deform spacetime.
\subsubsection{Explicit and spontaneous breaking}
A diffeomorphism transformation in GR has two equivalent
 interpretations called passive and active. Passive diffeomorphisms refer to coordinate transformations which act on the atlas 
 and leave objects invariant while altering the coordinate system. In contrast, active 
 diffeomorphisms are mappings from the manifold to itself, they move points within the manifold in the same coordinate system.
An active diffeomorphism affects
 dynamical tensor fields for geometry and matter,
 thus holding additional physical significance. One can say that passive 
 diffeomorphisms pertain to the formulation of the theory, whereas active diffeomorphisms relate to its dynamical 
 aspects. This distinction has been extensively examined in the literature~\cite{Rovelli}.
 
Under an active or passive diffeomorphism 
\begin{align}
x^{\mu}    \to x'^{\mu} (x)\,,
\end{align}
 the metric tensor transforms as
\begin{align}
   g'_{\mu \nu}(x')&= \frac {\partial x ^{\alpha}  }{\partial x'^{\mu}}   
    \frac {\partial x ^{\beta}  }{\partial x'^{\nu}}  g_{\alpha \beta}(x) \,.
\end{align}
This transformation rule
 applies to any dynamical tensor, however, when background fields are present, this equivalence may be 
 disrupted.
In such cases, it is useful to distinguish between observer and particle diffeomorphisms.
 An observer diffeomorphism is defined as a passive diffeomorphism, acting 
 covariantly on both dynamical objects and background fields. 
A particle diffeomorphism acts in an active manner, affecting only dynamical 
 objects while leaving the background fields invariant.
 
Furthermore, one distinguishes between explicit or spontaneous
 diffeomorphism 
breaking. An explicit background
 is introduced by hand in the effective Lagrangians and does not lead to additional degrees 
 of freedom compared to the unperturbed theory. In this case, the nondynamical background 
 has zero fluctuations and no associated equation of motion.
On the other hand, spontaneous background fields arise as vacuum solutions in an underlying
 theory. When the effective theory is truncated and written in terms of these backgrounds, an 
 equation of motion for the background field emerges, propagating additional degrees of 
 freedom.
 In this case one may have Nambu-Goldstone and massive modes~\cite{Bluhm:2004ep,Bluhm:2007bd}. 
 Explicit and spontaneous 
 backgrounds have been extensively discussed in the literature, and for further details, 
 we refer to the references~\cite{Bluhm:2023kph,Bluhm:2016dzm,Bluhm:2014oua,Bluhm:2007bd}.

 With these preparations, we are ready to discuss the potential conflict between dynamics and 
 geometry arising from nondynamical background fields. We focus on the generalized action
 containing explicit and spontaneous symmetry breaking
\begin{align}\label{Action_princ}
	S=\int  \mathrm{d}^4x  \sqrt{-g}  \left(    \prescript{(4)}{ }{R}+  \mathcal L_{1}(g_{\mu \nu}, \bar k^{\mu_1 \cdots \mu_n})   
	+ { \mathcal L_2}( g_{\mu \nu},\widetilde k^{\nu_1 \cdots \nu_m} ) \right)+ S_{\partial M}  \,.
\end{align}
The first term is the Einstein-Hilbert (EH) action, and we use the notation
 $ \prescript{(4)}{ }{R}$ for the scalar curvature in four-dimensional spacetime.
The Lagrangian $\mathcal{L}_1$ includes nondynamical 
background fields $\bar{k}^{\mu_1 \cdots \mu_n}$, while $\mathcal{L}_2$ involves dynamical background
 fields $\widetilde{k}^{\nu_1 \cdots \nu_m}$. 
 We choose $\mathcal L_1$ and $\mathcal L_2$ to be quadratic in 
 the background fields and to have contravariant indices, defining a specific diffeomorphism-violating theory.
 The boundary term $S_{\partial M} $ generally
 depends on the extrinsic curvature coupled to background fields and 
 ensures a proper Dirichlet variational formalism~\cite{Reyes:2021cpx, Reyes:2022mvm}.

An observer
 variation of the action with respect to the fields produces
\begin{align}\label{observer_t}
\delta S_{\rm {obs}}&= \frac 12 \int  \mathrm{d}^4x \sqrt{-g}  \Big[ \Big( G^{\mu \nu}   
+\left(T_1 \right)^{\mu \nu}  +(T_2)^{\mu \nu} \Big)   \delta g_{\mu \nu}+    \frac{\delta \mathcal L_1 }
{\delta \bar k^{\mu_1 \cdots \mu_n}} \delta \bar k^{\mu_1 \cdots \mu_n}   \notag \\
&\phantom{{}={}}\hspace{1.8cm}   +
\frac{\delta \mathcal L_2 }{\delta \widetilde k^{\nu_1 \cdots \nu_m}}   \delta \widetilde k^{\nu_1 \cdots \nu_m}  \Big ]  \,,
\end{align}
where we have the like energy-momentum tensor for explicit and spontaneous backgrounds 
 \begin{subequations}
\begin{align}
	(T_1)^{\mu \nu}&= \frac{1}{\sqrt{-g} }  \frac{\delta  (\sqrt{-g} \mathcal L_1)}{\delta g_{\mu \nu}} \,,
	 \\   (T_2)^{\mu \nu}&=\frac{1}{\sqrt{-g} }   \frac{\delta(  \sqrt{-g} \mathcal L_2)}{\delta g_{\mu \nu}} \,.
\end{align}
 \end{subequations}
Invariance under observer transformations $\delta S_{\rm {obs}}=0$ implies
 the modified Einstein's equation
\begin{align}\label{mod_E}
	G^{\mu \nu}+(T_{1})^{\mu \nu}  +(T_{2})^{\mu \nu}=0 \,,
\end{align}
and the equation of motion for the spontaneous background field 
\begin{align}
	\frac{\delta \mathcal L_2 }{\delta \widetilde k^{\mu_1 \cdots \mu_n}} \equiv \frac{\partial 
	\mathcal L_2 }{\partial \widetilde k^{\mu_1 \cdots \mu_n}}+\partial_{\lambda}
	\left( \frac{\partial \mathcal L_2 }{\partial  (\partial_{\lambda} \widetilde k^{\mu_1 \cdots \mu_n)}} \right) = 0\,.
\end{align}
Noteworthy, there is no equation of motion for the explicit background field.
Taking the covariant derivative of~\eqref{mod_E} we arrive at
\begin{align}\label{div}
	\nabla_{\mu} \left ( (T_{1})^{ \mu \nu}+  (T_{2}  ) ^{ \mu \nu} \right) =0 \,.
\end{align}
Now, consider an infinitesimal diffeomorphism transformation along a Killing field $\xi^{\mu}$
 \begin{align}
x'^{\mu}=x^{\mu}-\xi^{\mu} (x)\,,
\end{align}
which induces the variation in the tensors 
 \begin{subequations}\label{lie_derivatives}
\begin{align}
\delta g_{\mu \nu}=\nabla_{\mu} \xi_{\nu}+ 
\nabla_{\nu} \xi_{\mu} & \equiv \mathcal L_{\xi}  g_{\mu \nu}  \,,
\\
\delta \bar k^{\mu_1 \cdots \mu_n} &=\mathcal L_{\xi}  \bar k^{\mu_1 \cdots \mu_n} \,,
\\
\delta  \tilde k^{\nu_1 \cdots \nu_m}& =\mathcal L_{\xi}  \tilde k^{\nu_1 \cdots \nu_m}\,,
\end{align}
 \end{subequations}
 where $\mathcal L_{\xi}$ is the Lie derivative along the tensor field $\xi$ and of course 
 we have the isometry condition 
$\mathcal L_{\xi}  g_{\mu \nu} =0$.
 
Since the action~\eqref{Action_princ} is invariant under observer transformations, according 
to Noether theorem we will have a conserved current for the spontaneous field and a consistency 
condition for the explicit one. Considering ~\eqref{observer_t} and the isometry condition 
we arrive at
\begin{align}\label{obs_trans}
	\delta S_{\rm {obs}}&=\int  \mathrm{d}^4x 
	\partial_{\mu}   \Lambda^{\mu}  \,,
\end{align}
with 
\begin{align}\label{obs_trans}
	\Lambda^{\mu}= - \xi^{\mu} \sqrt{-g} ( \mathcal L_1+\mathcal L_2 ) \,,
\end{align}
where we have used the identity for a scalar density $\mathcal F$
of weight $1$
\begin{align}
	\mathcal L_{\xi}  \mathcal F &=\partial_{\mu}  (\xi^{\mu} \mathcal F) \,.
\end{align}

For a particle transformation, we arrive at
\begin{align}\label{part_t}
	\delta S_{\rm {part}}&=\int  \mathrm{d}^4x \left(  \sqrt{-g}
	 \big (G_{\mu \nu} 
	+(T_1)_{\mu \nu}  +(T_2)_{\mu \nu}  \big)     \delta g^{\mu \nu} +\sqrt{-g}
	\frac{\delta \mathcal L_2 }{\delta \widetilde k^{\nu_1 \cdots \nu_m}}   \delta \widetilde k^{\nu_1 \cdots \nu_m}   \right) \,.
\end{align}
We use the equations of motion to write 
\begin{align}
	 \frac{\partial \mathcal L_2 }{\partial \tilde k^{\mu_1 \cdots \mu_n}}=  \frac{1}{\sqrt{-g}} \partial_{\lambda} \left(\sqrt{-g}
	\frac{\partial \mathcal L_2 }{\partial  \partial_{\lambda} \widetilde k^{\mu_1 \cdots \mu_n}}  \right)\,.
\end{align}
Subtracting \eqref{obs_trans} with \eqref{part_t}, we have for the spontaneous sector
\begin{align}
	\delta S^{(2)}_{\rm{part}}- \int  \mathrm{d}^4x   \partial_{\mu} J_2^{\mu}  =\delta S^{(2)}_{\rm{obs}}   \,,
\end{align}
where we have defined the current
\begin{align}
	J_2^{\mu}  = \sqrt{-g}   \xi^{\lambda} \left(
	\frac{\partial \mathcal L_2 }{\partial ( \partial_{\lambda} \widetilde k^{\mu_1 \cdots \mu_n} )} 
	 -\delta^{\mu}  _{\lambda}  \mathcal L_2 \right) \equiv \xi^{\lambda} (T_2)^{\mu}_{\lambda} \,,
\end{align}
which is conserved due to 
\begin{align}
	\nabla_{\mu}  (T_2)^{\mu}_{\lambda}= 0 \,.
\end{align}
Hence, for spontaneous breaking one has $\delta S^{(2)}_{\rm part} =\delta S^{(2)}_{\rm obs}$ with no
  conflict between dynamics and geometry. We are defining $S^{(i)}=\int \mathrm{d}^4x \mathcal L_i$, with $i=1,2$.
However for the explicit background $\bar k^{\mu_1 \cdots \mu_n}$ one has
\begin{align}\label{condition_explicit}
	\delta S^{(1)}_{\rm{part}}+ \int  \mathrm{d}^4x  \sqrt{-g} \frac{\delta \mathcal L_1 }{\delta \bar k^{\mu_1 \cdots \mu_n}} 
	 \mathcal L_{\xi} \bar k^{\mu_1 \cdots \mu_n}=\delta S^{(1)}_{\rm{obs}}   \,,
\end{align}
where $\delta S^{(1)}_{\rm{obs}} =0$ and $\delta S^{(1)}_{\rm{part}} \neq 0$.
We see that the presence of the nondynamical background field $  \bar k^{\mu_1 \cdots \mu_n}$
leads to the
condition~\eqref{condition_explicit} that has to be imposed
on the explicit Lorentz violation in order to be consistent.
We can also write another condition to be verified for the background fields.
From~\eqref{part_t}, replacing the Lie derivative acting on the background,
 integrating by parts and neglecting a boundary term, we arrive at
\begin{align}
\nabla_{\mu} (  T_1    )^{\mu}_{\phantom{\mu}\nu}& =(J_1)_{\mu_1 \cdots  \mu_n}\nabla_{\nu}
{\bar k}^{\mu_1 \cdots  \mu_n}+ \nabla_{\mu_1} (   {\bar k}^{\mu_1 \cdots  \mu_n}  (J_1)_{\nu \mu_2 \cdots  \mu_n} )   +
\dots  \notag  \\ &\phantom{{}={}}\hspace{1.0cm}   \dots + \nabla_{\mu_n} ( {\bar k}^{\mu_1 \cdots  \mu_n}
(J_1)_{\mu_1 \cdots \mu_{n-1} \nu}  ) \,,
\end{align}
where
\begin{align}
(J_1)_{\mu_1 \cdots  \mu_n}  :=\frac{\delta \mathcal L_1 }{\delta \bar k^{\mu_1 \cdots \mu_n}} \,.
\end{align}
\subsubsection{Maximally symmetric subspaces }\label{SeccioII:Subsection:C}
In this subsection, we focus on explicit diffeomorphism breaking in a cosmological 
setting, where assumptions about spacetime geometry are established from the outset with
 the cosmological principle, incorporating spatial isotropy and homogeneity symmetries into the
  theory. As a result, the original $10$ generators of diffeomorphisms are reduced to $6$ Killing vector fields, 
  which describe rotations and translations. 

Consider the action 
\begin{align}\label{cosmological_action}
	S=\int _{\mathcal M} \mathrm{d}^4x  \sqrt{-g}  \left(    \prescript{(4)}{ }{R}+{  \mathcal L}(\bar k)
	 \right)+S_{\partial \mathcal M} \,,
\end{align}
where we denote the four-dimensional manifold by $\mathcal M$
and
we consider ${  \mathcal L}(\bar k)$ to depend solely on the explicit breaking. The last term is a boundary term
that includes extended GHY boundary terms.
The action is invariant under six Killing vectors fields which are also Noether symmetries, i.e., three generators
of rotations 
 \begin{subequations}
\begin{align}
\eta_{i}  &=\epsilon ^{ijk}  x^j \partial_k      \,,
\end{align}
and three generators of translations
\begin{align}
	\xi_i&=\left(\frac{k  x^i x ^j}{2}+\delta^{ij} \left( 1-\frac{k r^2}{4} \right)\right)   
   \partial_j \,,
\end{align}
 \end{subequations}
 with $\partial_i=\frac{\partial }{\partial \bar x^i}$ and $i=1,2,3$. We are using pseudo-cartesian coordinates $(t,\bar x,\bar y,\bar z)$
 where $\bar x=\bar r \sin \theta \cos \phi$, $\bar y=\bar r \sin \theta \sin \phi$, and $\bar z=\bar r \cos \theta$ and  the metric reads
 \begin{align}
   \mathrm{d}s^2=-\mathrm{d}t^2 + \frac{ a^2(t)  }{(1+\frac{k \bar r^2}{4})^2}   \Bigg( \mathrm{d}\bar x^2  +\mathrm{d}\bar y^2 +\mathrm{d}\bar z^2 
       \Bigg )  \,,
\end{align}
 with 
  \begin{align}
   \bar r=\frac{2r}{1+(1-kr^2)^{1/2}}  \,,
\end{align}
where $r$ correspond to the radial coordinate in the standard Friedmann metric with coordinates $(t,r,\theta,\phi)$.
We denote the set of six Killing vector fields by $\chi_{A}$, with $\chi_{1,2,3}  =\eta_{1,2,3}$,
$\chi_{4}=\xi_1$, $\chi_{5}=\xi_2$, $\chi_{6}=\xi_3$ and $A=1,\dots , 6$.

From the perspective of symmetries, the theory can be divided into two distinct sectors. The first sector
 involves the full diffeomorphism symmetry sector including for instance time translations transformations, where the action is not invariant
  under either particle or observer diffeomorphism transformations.
 In this broken diffeomorphism symmetry region, the backgrounds do not contribute to~\eqref{condition_explicit}
  considering their compatibility with restricted geometry and geometric couplings that imposes them to vanish. The second sector is 
  a maximally symmetric subspace
   characterized by isotropy and homogeneity where the theory in invariant under transformations involving the six Killing vectors $\chi_A$.
From ~\eqref{condition_explicit}, we have the condition in terms of the six Killing vector fields $\chi$ 
\begin{align}
	\delta S_{\rm{part}, \chi}+ \int_{\mathcal M}  \mathrm{d}^4x  \sqrt{-g} \frac{\delta \mathcal L(\bar k) }{\delta \bar k^{\mu_1 \cdots \mu_n}} 
	 \mathcal L_{\chi} \bar k^{\mu_1 \cdots \mu_n}=\delta S_{\rm{obs}, \chi}   \,.
\end{align}
  The condition is satisfied if the background fields remain form-invariant under both observer and particle diffeomorphisms.
 We have that $ \mathcal L_{\chi} \bar k^{\mu_1 \cdots \mu_n}=0$ in the maximally symmetric subspace and obtain
 $\delta S_{\rm {part},\chi}= \delta S_{\rm {obs},\chi} =0$. In this scenario, the explicit breaking is consistent.
\subsubsection{Isotropic and homogeneity conditions}\label{SeccioII:Subsection:C}
In the following, we aim to determine the conditions that must be imposed on the background fields to 
preserve the isotropy and homogeneity of the theory. In standard cosmology, three time-dependent 
parameters--the scale factor, pressure, and energy density of matter--enter into the Friedmann equations. 
These equations remain consistent as long as certain conditions are imposed on the energy-momentum 
tensor. In standard cosmology, the perfect fluid is form-invariant in the directions of the Killing vector fields. 
We adopt a similar approach here and impose the same condition on the like 
energy-momentum tensor associated to the background fields.
Therefore, we impose the condition
\begin{align}
	\mathcal L_{\chi}  { \mathcal T}^{ab}  =0\,,
\end{align}
where ${ \mathcal T}^{ab} $ represents the spatially energy-momentum tensor associated to the backgrounds
and the Lie derivative is taken in the direction of the six Killing fields.
A general solution in maximally symmetric subspaces
can be given in
terms of form-invariant tensors, see the reference~\cite{Weinberg}.
We have the general solution
\begin{align}\label{extra_cond_iso_hom}
\mathcal T_{ab}=f(t)q_{ab}  \,,
\end{align}
where $f(t)$ is a function of time.
Therefore, in order to preserve 
 isotropy and homogeneity one has to impose the extra condition~\eqref{extra_cond_iso_hom}.

For a general tensor $\mathcal T_{ab}$, its irreducible decomposition can be written
in terms of an antisymmetric, a symmetric traceless and a
trace part, whose decomposition reads
\begin{align}
\mathcal T_{ab}=\mathcal T _{[ab]}+ \mathcal T_{\langle ab \rangle}+ \frac{q_{ab}}{3}  \mathcal T  \,,
\end{align}
where $\mathcal T _{[ab]}$ is the antisymmetric part, and we have
the traceless part given by
\begin{align}
\mathcal T_{\langle ab \rangle}= \mathcal T_{(ab)}- \frac{q_{ab}}{3}  \mathcal T  \,,
\end{align}
and the trace $\mathcal T=q^{ab}  \mathcal T_{ab}$.
Indeed, one can count the independent components: $3$ for the antisymmetric part, $5$ 
for the symmetric trace free part and $1$ for the trace, which sum up to $9$ independent components in total.

 The extra condition to impose on the symmetric
 energy-momentum tensor to preserve isotropy and homogeneity 
 according to~\eqref{extra_cond_iso_hom} is
\begin{align}
\mathcal T_{\langle ab \rangle}=0  \,, \qquad f(t)=\frac{\mathcal T}{3}\,,
\end{align}
which will be used in the next sections.
\section{Explicit background fields}\label{SectionIII}
\subsection{The t-sector of the minimal gravitational SME }\label{SectionIII:Sub:A}
Consider the Einstein-Hilbert (EH) action with cosmological constant $\Lambda$
plus the 
minimal gravitational SME action~\cite{Kostelecky:2003fs},
\begin{subequations}
\begin{align}\label{mod_action}
S&=S_{\mathrm{EH}}+ S_{\mathrm{ SME}}  \,,
\displaybreak[0]\\[2ex]  \label{modEH_action}
S_{\mathrm{EH}}&= \frac{1}{2\kappa}  \int_{\mathcal{M}} \mathrm{d}^4x \,
\sqrt{-g}  \left(   \prescript{(4)}{ }{R}-2\Lambda  \right) \,,
   \displaybreak[0]\\[2ex]   \label{eq:effective-action}
 S_{\mathrm{SME}}&=\frac{1}{2\kappa}  \int_{\mathcal{M}} \mathrm{d}^4x \,
\sqrt{-g}  \,   (  k_R)^{\mu \nu \rho \sigma} \,  \prescript{(4)}{ }{R}_{\mu \nu \rho \sigma}   \,,
\end{align}
\end{subequations}
with $\kappa= 8\pi G_N$,  and we denote the Riemann tensor
$ \prescript{(4)}{ }{R}_{\mu \nu \rho \sigma}$, the Ricci tensor $  \prescript{(4)}{ }{R}_{\mu\nu}:=
  \prescript{(4)}{ }{R}^{\alpha}_{\phantom{\alpha} \mu \alpha \nu}$  and the Ricci scalar
$\prescript{(4)}{ }{R}:=  \prescript{(4)}{ }{R}^{\mu}_{\phantom{\mu}\mu}$ 
of the four-dimensional spacetime 
manifold $\mathcal M$ with metric 
tensor $g_{\mu \nu}$ and $g=\rm{det} (g_{\mu \nu })$.
The explicit background field $(k_R)^{\mu \nu \rho \sigma}(x)$ is spacetime-dependent 
and inherit the symmetries of the Riemann tensor. 

The gravity SME action~\eqref{eq:effective-action} 
can be written in terms of independent components of the explicit background as
 \begin{align}\label{eq:explicit_diffeo}
 S_{\mathrm{SME}}&= \frac{1}{2\kappa}\int_{\mathcal{M}} \mathrm{d}^4x 
\sqrt{-g}   \left(-u\,    \prescript{(4)}{ }{R} + s^{\mu \nu}
 \prescript{(4)}{ }{R}_{\mu \nu}  
+t^{\mu \nu \rho \sigma} 
 \prescript{(4)}{ }{R}_{\mu \nu \rho \sigma}\right)\,.
\end{align}
The coefficients $t^{\mu \nu \rho \sigma}$ and $ s^{\mu \nu}$ enjoy
 the symmetries of the Riemann and Ricci tensor, respectively,
and $u$ is a scalar background field.
In the case of 
 spontaneous breaking with the occurrence of transmutation of 
 geometric tensors due to redefinitions, it is more
  convenient to consider the irreducible decomposition of
 $(k^{(4)}_R)^{\mu \nu \rho \sigma}$ yielding
 \begin{align}\label{S_LV_spontaneous}
(  k^{(4)}_R)^{\mu \nu \rho \sigma} \, \prescript{(4)}{ }{R}_{\mu \nu \rho \sigma} &=     -u    \prescript{(4)}{ }{R}   + s^{\mu \nu}\,
 \prescript{(4)}{ }{R}^T_{\mu \nu}   + t^{\mu \nu \rho \sigma} 
  \prescript{(4)}{ }{C}_{\mu \nu \rho \sigma}    \,,
\end{align}
where $\prescript{(4)}{ }{C}_{\mu \nu \rho \sigma} $ is   
the four dimensional Weyl tensor, and $ \prescript{(4)}{ }{R}^T_{\mu \nu}$
is the traceless Ricci tensor. Indeed, in this case the $20$ independent components are 
distributed as follows: $1$ 
for $u$, $9$ for the traceless and symmetric background $ s^{\mu \nu}$, and $10$ for $t^{\mu \nu \rho \sigma}$.

Let us focus on the $t$-sector 
of the action~\eqref{eq:explicit_diffeo} by setting
 $u=0$ and $s^{\mu \nu}=0$, such that our gravity
  model is described by the action 
\begin{align}\label{eq:action-initial}
S_G&=\int_{\mathcal{M}} \mathrm{d}^4x   \frac{    \sqrt{-g} }{2\kappa}   \left( \prescript{(4)}{ }{R}-2\Lambda
+t^{\mu \nu \rho \sigma} \prescript{(4)}{ }{R}_{\mu \nu \rho \sigma}   \right) \,.
\end{align}

The modified Einstein's equation for~\eqref{eq:action-initial} 
have been derived in~\cite{Bailey:2006fd} and read 
\begin{align}\label{Eq:generaldecompositiont}
    G^{\mu\nu}+\Lambda g^{\mu\nu}-\left(T^{Rt}\right)^{\mu\nu}
    &=0 \,,
\end{align}
with $G^{\mu\nu}$ the Einstein tensor and 
 $\left( T^{Rt}\right)^{\mu\nu}$ is given by
\begin{align}\label{Eq:equationst}
(T^{Rt})^{\mu\nu}&= \frac{1}{2}t^{\alpha\beta\gamma\mu} \prescript{(4)}{ }{R}_{\alpha\beta\gamma}^{
\phantom{\alpha} \phantom{\beta} \phantom{\gamma}\nu}+\frac{1}{2}
t^{\alpha\beta\gamma\nu}   \prescript{(4)}{ }{R}_{\alpha\beta\gamma}^{\phantom{\alpha} 
\phantom{\beta} \phantom{\gamma} \mu} 
+\frac{1}{2}   g^{\mu\nu}  t^{\alpha\beta\gamma\delta}
\prescript{(4)}{ }{R}_{\alpha\beta\gamma\delta}  \nonumber    \\ 
&\phantom{{}={}}\hspace{0.5cm}          -\nabla_\alpha\nabla_\beta t^{\mu\alpha\nu\beta}
-\nabla_\alpha\nabla_\beta t^{\nu\alpha\mu\beta} \,.
\end{align}

According to the ADM formalism
we decompose the background 
$t^{\mu \nu \rho \sigma}$ in its normal and tangent components to the hypersurface $\Sigma_t$.
For this we consider the delta in Eq.~\eqref{comp_delta}
and write
\begin{align}\label{previouspdecomposition}
t^{\alpha\beta\gamma\delta}=\left(e^\alpha_{a} \widetilde E_\beta^{ a}-n^\alpha n_\mu \right)  
  \left(e^\beta_{b} \widetilde E_\nu^{ b}  -n^\beta n_\nu \right)
    \left(e^\gamma_{c} \widetilde E_\rho^{ c }-n^\gamma n_\rho \right)  \left(e^\delta_{d} \widetilde 
    E_\sigma^{ d}-n^\delta n_\sigma \right)      t^{\mu\nu\rho\sigma}   \,.
\end{align}
We arrive at
\begin{align}\label{pdecomposition}
t^{\alpha\beta\gamma\delta}&=    ( e^\alpha_{a}e^\beta_{ b}e^\gamma_{ c}e^\delta_{d}  )
t^{abcd} + \left(  \mathcal P^{\alpha\beta\gamma\delta}_ {bcd} \right) t^{\mathbf{n}  bcd} +\left(\mathcal 
Q^{\alpha\beta\gamma\delta}_ {bd} \right)t^{\mathbf{n}  b \mathbf{n} d}\,,
\end{align}
with
\begin{subequations}
\begin{align}\label{eq:proj1}
\mathcal P^{\alpha\beta\gamma\delta}_ {bcd}= -n^\alpha e^\beta_{ b}e^\gamma_{c}e^\delta_{d} 
    +e^\alpha_{b}n^\beta e^\gamma_{ c}e^\delta_{d}    -e^\alpha_{c}e^\beta_{d}n^\gamma e^\delta_{b}+e^\alpha_{c}
    e^\beta_{d}e^\gamma_{b}n^\delta   \,,
\end{align}
\begin{align}\label{eq:proj2}
\mathcal Q^{\alpha\beta\gamma\delta}_ {bd}&=n^\alpha e^\beta_{b}n^\gamma e^\delta_{d}-n^\alpha e^\beta_{b}e^\gamma_{d}n^\delta
     -e^\alpha_{b}n^\beta n^\gamma e^\delta_{d}     +e^\alpha_{ b}n^\beta e^\gamma_{d}n^\delta \,.
\end{align}
\end{subequations}
Above, we have defined  $t^{\mathbf{n}  
 bcd}:=n_{\alpha}\widetilde E_\nu^b \widetilde E_\rho^d \widetilde E_\sigma^d t^{\alpha \nu\rho\sigma}$
 and 
 $ t^{\mathbf{n}  
 b\mathbf{n}d}:=n_{\alpha}\widetilde E_\rho^b n_\beta \widetilde E_\sigma^d t^{\alpha \rho\beta\sigma}$ 
and according to the symmetries
 of $t^{\alpha\beta\gamma\delta}$ we have neglected third and fourth contractions with $n_{\mu}$.

Consider the Gauss-Codazzi equations 
\begin{subequations}\label{GC}
\begin{align}
    e^\alpha_{a}e^\beta_{b}e^\gamma_{ c}e^{\delta}_{d}  \prescript{(4)}{ }{R}_{\alpha\beta\gamma\delta}&= R_{abcd}-\left(K_{ad}K_{bc}-K_{ac}K_{bd}\right)  \\
   n^\alpha e^\beta_{b}e^\gamma_{c}e^{\delta}_{d}\prescript{(4)}{ }{R}_{\alpha\beta\gamma\delta}&= D_d K_{bc}-D_c K_{bd} \\
   n^\alpha e^\beta_{b}n^\gamma e^{\delta}_{d} \prescript{(4)}{ }{R}_{\alpha\beta\gamma\delta}&=\frac{1}{N}D_b D_d N   + K^{ c}_{d}K_{b c} 
    - \frac{1}{N} e^{\rho}_{b}e^{\sigma}_{d}\mathcal{L}_m \left(\widetilde E_{\rho}^{e}\widetilde E_{\sigma}^{f}K_{ef}   \right) \,,
\end{align}
\end{subequations}
where $R_{abcd}$ and $K_{ab}$ are the induced intrinsic and extrinsic curvatures, respectively,
on the hypersurface $\Sigma_t$ and $D_{a}$ is the covariant derivative metric 
compatible with the induced metric $q_{ab}$. We also have introduced the four vector $m^{\mu}=Nn^{\mu}$
and the Lie derivative along $m^{\mu}$, $\mathcal{L}_m K_{\nu\sigma}=\dot K_{\nu\sigma}- \mathcal{L}_{N} K_{\nu\sigma} $
where the dot means derivative with respect to $t$.

Using the decomposition of the background in Eq.~\eqref{pdecomposition}, we contract with the 
Riemann tensor by using Eqs.~\eqref{GC} finding three sectors of the decomposition 
\begin{align}\label{eq:tandR}
 t^{\alpha\beta\gamma\delta}  \prescript{(4)}{ }{R}_{\alpha\beta\gamma\delta}&= t^{abcd}    {  \mathcal{R}}_{abcd}
-4   t^{\mathbf{n}abc}   {  \mathcal{R}}_{\mathbf{n} abc}
+4    t^{\mathbf{n}a\mathbf{n}b}   {  \mathcal{R}}_{\mathbf{n} a\mathbf{n} b} \,,
\end{align}
with
\begin{subequations}\label{eqs:Rdef}
\begin{align}
  \mathcal{R}_{abcd}&:=R_{abcd}-(K_{ad}K_{bc}-K_{ac}K_{bd}) \,,  \\
   \mathcal{R}_{\mathbf{n}abc}&:=D_b K_{ac}-D_c K_{ab}   \,, \\
 \mathcal{R}_{\mathbf{n}a\mathbf{n}b}&:=\frac{1}{N}D_a D_b N   + K^{ c}_{b}K_{a c} 
    - \frac{1}{N} e^{\rho}_{a}e^{\sigma}_{b}\mathcal{L}_m (\widetilde E_{\rho}^{c}\widetilde E_{\sigma}^{d}K_{cd})\,.
\end{align}
\end{subequations}
Everything in the right hand side in Eq.~\eqref{eq:tandR} has been written with 
space indices by 
using both expressions in Eqs.~\eqref{eq:indices_hypersurface}.
We also note that a similar theory involving auxiliary fields has been studied in the reference~\cite{Deruelle:2009zk}.

The decomposition of the EH action in~\eqref{eq:action-initial} follows by considering
\begin{align}\label{EH_decom}
   \prescript{(4)}{ }{R}&= R -K^2+K_{ab}K^{ab}-2\nabla_{\mu}\zeta^{\mu}   \,,
\end{align}
where $ R$ is the induced three curvature on the hypersurface,
and we have defined $ \zeta^{\mu}:=a^{\mu}-Kn^{\mu}$
in terms of the acceleration $a^{\mu}:=n^{\nu}\nabla_{\nu}n^{\mu}$.

We can write the 
action~\eqref{eq:action-initial} in the form 
\begin{align}\label{ADM_EH_mod}
S _G
 &=\int_{\mathcal{M}} \mathrm{d}^3y \frac{N\sqrt{q}}{2\kappa} 
    \mathrm{d}t  \left( \mathcal{L}_{\mathrm{EH}}+ \mathcal{L}^{(1)}+\mathcal{L}^{(2)}+\mathcal{L}^{(3)}  \right)  \,.
\end{align}
We have the usual EH term  
\begin{subequations}
\begin{align}\label{EH_ADM_0}
    \mathcal{L}_{\mathrm{EH}}&=\bigg( R-2\Lambda -K^2+K_{ab}K^{ab}-2\nabla_{\mu}\zeta^{\mu}  \bigg)  \,,
\end{align}
a purely spatial sector
\begin{align}
\mathcal{L}^{(1)}&=   t^{abcd} \bigg( R_{abcd}+K_{ac}K_{bd}-K_{ad}K_{bc}   \bigg)\,,
\end{align}
a normal sector
 \begin{align}
    \mathcal{L}^{(2)}&=- 4t^{\mathbf{n}abc} \bigg( D_b K_{ac}-D_c K_{ab}    \bigg)\,,
\end{align}
   and a doubly-normal sector
\begin{align}  \label{Lag3}
    \mathcal{L}^{(3)}&=  4 t^{\mathbf{n}a  \mathbf{n}b} \bigg(   -\frac{1}{N} e^\rho_{a}e^\sigma_{b} \mathcal{L}_m 
        \big(\widetilde{E}_\rho^{c}\widetilde{E}_\sigma^{d}K_{cd}\big)+\frac{1}{N}D_a D_b N   + K^c_{\phantom{c} a}K_{c b}   \bigg)    \,,
\end{align}
\end{subequations}
where we have used $\sqrt{-g}=N\sqrt{q}$.
\subsection{Extended boundary terms}\label{SectionIII:subC}
As is well known, the EH action \eqref{ADM_EH_mod} includes terms involving linear accelerations, whose variations are non-zero in the normal directions. We establish here the correct variational formalism that incorporates these boundary terms. This is explicitly manifested in the EH term \eqref{EH_ADM_0}, and additionally, we introduce a boundary term arising from the modified sector \eqref{Lag3}. To handle these aspects, we employ the standard formalism of surface terms in General Relativity, extending from the Gibbons-Hawking-York (GHY) terms. Below we generalize
previous studies in the $u$ and $s^{\mu \nu}$ sectors of the SME~\cite{Reyes:2021cpx,Reyes:2022mvm,Reyes:2023sgk} to the case including the $t^{\mu \nu \rho \sigma}$
background.

We start by
removing the second derivative terms in~\eqref{Lag3} and by using the identity
\begin{align}\label{eq:derivativehd_term}
  \frac{1}{N}t^{\mathbf{n}a\mathbf{n}b}  e^{\rho}_{a} e^{\sigma}_{b}\mathcal{L}_m (\widetilde E_{\rho}^{c}\widetilde E_{\sigma}^{d}K_{cd})
    &=\nabla_\lambda (n^\lambda t^{\mathbf{n}a\mathbf{n}b} K_{ab}) -\frac{1}{N}K_{cd}\widetilde E_{\rho}^{c}\widetilde E_{\sigma}^{d}\mathcal{L}_m \big(e^{\rho}_{a}
     e^{\sigma}_{b} t^{\mathbf{n}a\mathbf{n}b}\big)    \nonumber \\
     &-K K_{ab} t^{\mathbf{n}a\mathbf{n}b}\,.
\end{align}
From~\eqref{EH_ADM_0} and ~\eqref{Lag3} it follows  
the standard GHY term and an extended GHY boundary
 term, which will be added conveniently from the start in order to avoid 
linear higher derivatives in the action.
We can write the ADM action in terms of 
standard and extended GHY boundary terms as 
\begin{align}
 S_{\mathrm{ADM}}&=\frac{1}{2\kappa}\int_{\mathcal{M}} 
 \mathrm{d}t \mathrm{d}^3y\,N\sqrt{q}\Big({ \mathcal L}_0+ 
{\mathcal L}'  \Big)+S_{   \partial \mathcal{M}  }    ^{({t})}  \,,
\end{align}
where 
\begin{align}
{ \mathcal L}_0&=  R -K^2+K_{ab}K^{ab} -2\Lambda  \,,
\\
{\mathcal L}'&=   t^{abcd}     {  \mathcal{R}}_{abcd}-4
t^{\mathbf{n}abc}  {  \mathcal{R}}_{\mathbf{n} abc} +     4 t^{\mathbf{n}a\mathbf{n}b}    {  \mathcal{R}}'_{\mathbf{n} a \mathbf{n}b}     
+\frac{1}{N} K_{cd}\widetilde E_{\rho}^{c}\widetilde E_{\sigma}^{d}\mathcal{L}_m \big(e^{\rho}_{a}
     e^{\sigma}_{b} t^{\mathbf{n}a\mathbf{n}b}\big)\,,
\end{align}
and we have defined
\begin{align}
  {  \mathcal{R}}'_{\mathbf{n} a \mathbf{n}b} =      \bigg(KK_{ab}  
+ \frac{1}{N}D_a D_b N + K^c_{\phantom{c} a}K_{c b} \bigg)   \,.
\end{align}
The generalization of the  GHY boundary term is given by
\begin{align}
S_{   \partial \mathcal{M}  }    ^{({t})}   =\frac{1}{2\kappa} \oint_{\partial\mathcal{M}}  
 \mathrm{d}^3y \, \varepsilon \sqrt{q} \,    \left(2K+ 4 t^{\mathbf{n}a\mathbf{n}b} K_{ab}  \right)  \,,
\end{align}
where $\varepsilon=n_{\mu} n^{\mu}$ is evaluated in each point of $\partial\mathcal{M}$, which can be
timelike ($\varepsilon=1$), spacelike ($\varepsilon=-1$) or lightlike ($\varepsilon=0$).
\subsection{Modified Friedmann equations}\label{SectionIV}
Consider the Friedmann-Lema\^{i}tre-Robertson-Walker (FLRW) metric
\begin{align}
   \mathrm{d}s^2=-\mathrm{d}t^2 + a^2(t) \left[ \frac{\mathrm{d}r^2}{1-kr^2}
    +r^2  \left( \mathrm{d}\theta^2+\sin^2\theta \mathrm{d}\phi^2     \right)  \right ]  \,,
\end{align}
with the cosmic scale factor $a(t)$, and the spatial part 
written in terms of three-dimensional spherical coordinates $(r,\theta,\phi)$. Here $k$
is the spatial curvature that describes a closed ($k=1$), flat ($k=0$) and open ($k=-1$) universe.
We consider a perfect fluid composition with
energy-momentum tensor  
\begin{align}\label{E-M_perfectfluid}
    (T_m)^{\mu\nu}=(\rho+P)U^\mu U^\nu +Pg^{\mu\nu}\,,
\end{align}
where $\rho$ is the fluid density, $P$ its pressure, and $U^\mu$ its four-velocity. 
It is common to consider a fluid at the rest frame, such that $U^{\mu}=(1,0,0,0)$. 

Variation of the 
action~\eqref{eq:action-initial} with a matter source $(T_m)^{\mu\nu}$,
gives
 the modified Einstein's equation
\begin{align}
    G^{\mu\nu}-\big(T^{Rt}\big)^{\mu\nu}
    &=\kappa(T_m)^{\mu\nu} \,,
\end{align}
where we have set the cosmological term to zero.
We find the standard decomposed Einstein tensor
\begin{align}
G^{00}&=    3\left(\frac{k}{a^2(t)}+H^2  \right)    \,,
\\
G^{ab}&=     -\left(\frac{k}{a^2(t)}+2  \left(\dot{H}+H^2 \right)+H^2\right)q^{ab}    \,.
\end{align}

The decomposed like energy-momentum tensor for the $t$-background
can be written in the compact form
\begin{align}\label{energy-mom_tensor_tsector}
    (T^{Rt})^{\mu\nu}&= e^\mu_{a}e^\nu_{b}   ( T_1)^{ab}+ \left(e^\mu_{a}n^\nu  +n^\mu e^\nu_{a}  \right) ( T_2)^a
+n^\mu n^\nu    T_3\,,
\end{align}
where we have defined
the purely spacelike component
\begin{align}
 ( T_1)^{ab}&= 2\left( \frac{k}{a^2(t)}+H^2\right)  q_{cd}t^{acbd}+q^{ab}\left[   \bigg(\frac{k}{a^2(t)}+H^2\bigg)  q_{cd}
   q_{rs}t^{crds} -2\big(H^2  + \dot{H} \big)q_{cd}t^{\mathbf{n}c \mathbf{n}d}\right]   \nonumber \\
    &-\Big( D_c D_d t^{acbd}+ D_c D_d t^{bcad}+2  \ddot{t}^{a\mathbf{n}b\mathbf{n}}    - 
     2 D_c\dot{t}^{acb\mathbf{n}}-2 D_c\dot{t}^{bca\mathbf{n}}+2H q_{cd}\dot{t}^{acbd}+ 16H \dot{t}^{a\mathbf{n}b\mathbf{n}} \nonumber
       \\&-10HD_c t^{a\mathbf{n}bc} -10HD_c t^{b\mathbf{n}ac}+2(\dot{H}+6H^2)q_{cd}  t^{acbd} + 2(5\dot{H}+17H^2)t^{a  \mathbf{n}b \mathbf{n}} \Big)  \,,
\end{align}  
the mixed component
\begin{align}\label{mom_constraint}
   (  T_2)^a&=    D_c D_dt^{ac\mathbf{n}d} +D_c D_d t^{\mathbf{n}cad}- D_c\dot{t}^{\mathbf{n}ca\mathbf{n}} -2 H  q_{cd}D_e t^{acde}   
-5HD_c  t^{a\mathbf{n}\mathbf{n}c}   \nonumber \\
&\phantom{{}={}}\hspace{1.0cm} +H  q_{cd} \dot{t}^{acd\mathbf{n}}-\bigg( \dot{H}+11H^2+\frac{k}{a^2(t)}\bigg) q_{cd}t^{cad\mathbf{n}}   \,,
\end{align}
and the doubly-normal component
\begin{align}
 T_3&=   \bigg(\frac{k}{a^2(t)}+H^2\bigg) q_{ab} q_{cd}t^{acbd}-2\Big( D_c D_dt^{\mathbf{n}c\mathbf{n}d} 
+H q_{cd}\dot{t}^{\mathbf{n}cd\mathbf{n}}- 2H q_{cd}D_at^{ca\mathbf{n}d}  \nonumber \\
&\phantom{{}={}}\hspace{1.0cm} +H^2h_{cb} q_{ad}t^{acbd}-2(\dot{H}+3H^2) q_{cd}t^{\mathbf{n}c\mathbf{n}d}\Big) \,.
\end{align}
Recall in normal coordinates the ones we use in the context of cosmology one has $e^{\mu}_a=\delta ^{\mu}_a$
and $n^{\mu}=(1,0,0,0)$.
After a lengthly calculation the first Friedman equation reads
\begin{align}\label{FEq_Friedmann}
        H^2&=\frac{1}{3\Big(1-\frac{1}{3}q_{ab}q_{cd}t^{acbd}-2q_{ab}t^{a\mathbf{n}b\mathbf{n}}\Big)}\bigg[\kappa \rho -\frac{k}{a(t)^2}q_{ab}q_{cd}t^{acbd}  -2\Big( D_a D_bt^{\mathbf{n}a\mathbf{n}b}     \\
      &-Hq_{ab}\dot{t}^{\mathbf{n}a\mathbf{n}b}    - 2Hq_{cb}D_at^{ca\mathbf{n}b}-H^2q_{ab}
       q_{cd}t^{acbd}-\big(2\dot{H}+3H^2\big)h_{ab}t^{\mathbf{n}a\mathbf{n}b}\Big)-\frac{3k}{a(t)^2}\bigg] \,,  \nonumber 
\end{align}
and the second Friedmann equation can be written as
 \begin{align}
 &   \dot{H}+H^2=\frac{1}{(-6)\Big(1-\frac{1}{3}q_{ab}q_{cd}t^{acbd}-2q_{ab}t^{a\mathbf{n}b\mathbf{n}}\Big)}\bigg[\kappa (\rho+3P) 
 +4\bigg(\frac{k}{a(t)^2}+H^2\bigg)q_{ab}q_{cd}t^{acbd}   \nonumber \\
    & -2\Big( D_c D_d (q_{ab}t^{acbd})+ D_a D_bt^{\mathbf{n}a\mathbf{n}b}+  q_{ab}\ddot{t}^{a\mathbf{n}b\mathbf{n}}
    -2 D_c(q_{ab}\dot{t}^{acb\mathbf{n}})+Hq_{ab}q_{cd}\dot{t}^{acbd}+ 7H q_{ab}\dot{t}^{a\mathbf{n}b\mathbf{n}}  \nonumber \\
    &-8HD_c (q_{ab}t^{a\mathbf{n}bc}) +4H^2q_{ab}q_{cd}t^{acbd}+ 8H^2q_{ab}t^{a\mathbf{n}b\mathbf{n}}\Big) \bigg] \,.
 \end{align}
For details on the calculation see the Appendix~\ref{AppendixB}.
We also find the traceless symmetric part of the like energy-momentum tensor associated to the background
to be
\begin{align}\label{extra_condition_t}
 T^{\langle ab \rangle}=& -2\bigg(\frac{k}{a(t)^2}-(\dot{H}+5H^2)\bigg)\Big(h_{cd}t^{acbd}-\frac{1}{3}q^{ab}q_{cd}q_{ef}t^{cedf}\Big)
 +2(5\dot{H}+17H^2)\Big(t^{a\mathbf{n}b\mathbf{n}}    -\frac{1}{3}q^{ab}q_{cd}t^{c\mathbf{n}d\mathbf{n}}\Big)\nonumber \\
   &+2  \Big(\big(\ddot{t}^{a\mathbf{n}b\mathbf{n}}+Hq_{cd}\dot{t}^{acbd}+ 8H \dot{t}^{a\mathbf{n}b\mathbf{n}}\big)-\frac{1}{3}q^{ab}
   q_{ef}\big(\ddot{t}^{e\mathbf{n}f\mathbf{n}}+Hq_{cd}\dot{t}^{cedf}+ 8H \dot{t}^{e\mathbf{n}f\mathbf{n}}\big)\Big) \nonumber \\
    &+\Big( D_c 
    D_d t^{acbd}+ D_c D_d t^{bcad}-2 D_c\dot{t}^{acb\mathbf{n}}-2 D_c\dot{t}^{bca\mathbf{n}}-10HD_c t^{a\mathbf{n}bc} 
    -10HD_c t^{acb\mathbf{n}}  \Big)  \notag \\ & -\frac{1}{3}q^{ab}q_{ef}\Big( 2D_c 
    D_d t^{ecfd}-4 D_c
    \dot{t}^{ecf\mathbf{n}}   
     -20HD_c t^{e\mathbf{n}fc} \Big) \,.
 \end{align}
\subsection{Late acceleration}\label{SectionV}
We introduce a simplification 
and focus on the purely spatial sector of $t^{abcd}$.
Our aim is to determine if this specific example can account for the accelerated 
expansion of the universe using only standard matter and radiation. For this particular
 case, the tensor in equation \eqref{energy-mom_tensor_tsector} simplifies to
\begin{align}\label{TRtspatial}
  (T_{\textrm{t-spacelike}}^{Rt})^{\mu\nu}&= e^\mu_{a}e^\nu_{b}\bigg[ 2\bigg(\frac{k}{a(t)^2}+H^2\bigg)
    q_{cd}t^{acbd}    +\bigg(\frac{k}{a(t)^2}+H^2\bigg)q^{ab}q_{cd}q_{ef}t^{cedf} \nonumber \\
    & -\Big( D_c D_d t^{acbd}+ D_c D_d t^{bcad}    +2Hq_{cd}\dot{t}^{acbd}
    +2(\dot{H}+6H^2)q_{cd}t^{acbd}\Big)\bigg] \nonumber \\
&- e^\mu_{ a}n^\nu\bigg[  2 Hq_{bc}D_d t^{acbd}\bigg] -n^\mu e^\nu_{b}\bigg[ 2Hq_{ad}D_ct^{acbd}\bigg]  \notag \\
&+n^\mu n^\nu\bigg[\bigg(\frac{k}{a(t)^2}+3H^2\bigg)q_{ab}q_{cd}t^{acbd}\bigg] \, .
\end{align}
Also, from \eqref{FEq_Friedmann} the first Friedmann equation can be written as
\begin{align}
H^2 &=\frac{1}{3\Big(1-\frac{1}{3}q_{ab} q_{cd}  t^{acbd}\Big)}  \bigg( \kappa \rho -\frac{k}{a(t)^2}
q_{ab}q_{cd}t^{acbd}    -\frac{3k}{a(t)^2}\bigg)  \,,
\end{align}
and the second Friedmann equation as
\begin{eqnarray}
\dot{H}+H^2&=&\frac{1}{(-6)\Big(1-\frac{1}{3}q_{ab}q_{cd}t^{acbd}\Big)}\bigg[\kappa (\rho+3P) 
+\frac{4k}{a(t)^2}q_{ab}q_{cd} t^{acbd}     -2q_{ac}\Big[ D_b D_d t^{abcd}\notag \\ &&
+Hq_{bd}\dot{t}^{abcd}   +2H^2q_{bd}t^{abcd}\Big] \bigg]  \,.
 \end{eqnarray}
The traceless symmetric tensor in~\eqref{extra_condition_t} reduces to
\begin{align}
\label{tperp0}
 T_{\textrm{t-spacelike}}^{\langle ab \rangle}&= 2\bigg(\frac{k}{a(t)^2}-(\dot{H}+5H^2)\bigg)  \bigg(q_{cd}t^{acbd}-\frac{1}{3}q^{ab}q_{ef}q_{cd}t^{ecfd}\bigg) 
 -2H\Big(q_{cd}\dot{t}^{acbd}-\frac{1}{3}q^{ab}q_{ef}q_{cd}\dot{t}^{ecfd}\Big) \nonumber \\
    &-\bigg(\Big( D_c D_d t^{acbd}+ D_c D_d t^{bcad}\Big)-\frac{1}{3}q^{ab}q_{ef}\Big( D_c D_d t^{ecfd}+ D_c D_d t^{fced}   \Big)\bigg)\,.
\end{align}
Since we are interested in the preservation of isotropy and homogeneity 
we impose the extra condition $ T_{\textrm{t-spacelike}}^{\langle ab \rangle}=0$. For this, we will assume
\begin{eqnarray}\label{rel1}
 q_{cd}t^{acbd}-\frac{1}{3}q^{ab}q_{ef}q_{cd}t^{ecfd}&=&0 \,.
  \end{eqnarray} 
This choice results to be consistent and it also determines the form 
of $t$. We left the details in Appendix \ref{AppendixC} where we show that the tensor $t$ possesses the following form
\begin{eqnarray}\label{explicit_sol}
   t^{r\phi\theta\phi}&=&t^{r\theta\phi\theta}=t^{\theta r\phi r}=0\,, \\
  t^{r\theta r \theta} &=&   \frac{ (1-kr^2)\eta(t)}{r^2} \,,\\
     t^{r\phi r\phi}&=& \frac{ (1-kr^2)\eta(t)}{r^2 \sin^2\theta } \,, \\
t^{\theta\phi\theta\phi}&=&\frac{ \eta(t)}{r^4 \sin^2\theta } \,.
\end{eqnarray}
In order to take the simplest example to our analysis, we consider $\eta$ to be constant.
With these solutions we write the Friedmann equations considering the current observations of $k=0$ \cite{Planck:2018vyg}, arriving at  
\begin{eqnarray}
H^2 &=& \frac{\kappa \rho}{3 (1 - 2 a^4 \eta)}\,, \label{frieda}\\
\dot{H} + H^2 &=& - \frac{1}{6 (1 - 2 a^4 \eta)} [\kappa (\rho + 3 P)   - 24 a^4
H^2 \eta] \,. \label{friedb}
\end{eqnarray}
We assume standard matter, and use an equation of state defined by 
$P = w \rho$, where $w=0$ is for non-relativistic matter and $w=1/3$ for radiation. Also, we will use the definition of the deceleration parameter
\begin{equation}
q = - \frac{\ddot{a}a}{\dot{a}^2}\,.
\end{equation}
With these definitions, the second Friedman equation now reads
\begin{equation}
\label{friedb2}
H^2 = \frac{1 + 3 w}{6 (q (1 - 2 \eta a^4) + 4 \eta a^4)} \rho\,.
\end{equation}

In this way, we only need to solve the first Friedmann equation to get the solution, and the expression \eqref{friedb2} gives the functional form for the decceleration parameter. Thus, for consistency we equate \eqref{frieda} with \eqref{friedb2} and we obtain that
\begin{equation}
\label{qeq}
q = \frac{(1 + 3 w) - 2 (5 + 3 w)\eta a^4}{2 (1 - 2 \eta a^4)}\,.
\end{equation}

We have an accelerated expansion when $q<0$, and this is satisfied when $1 - 2 \eta a^4 > 0$ and $(1 + 3 w) - 2 (5 + 3 w) \eta a^4 < 0$, which can be summarized as
\begin{equation}
\label{conditionsacc}
\left( \frac{1 + 3 w}{5 + 3 w} \right) \frac{1}{2 \eta} < a^4 < \frac{1}{2 \eta}\,.
\end{equation}

\begin{figure}[h]
	\centering
	\includegraphics[scale=0.4]{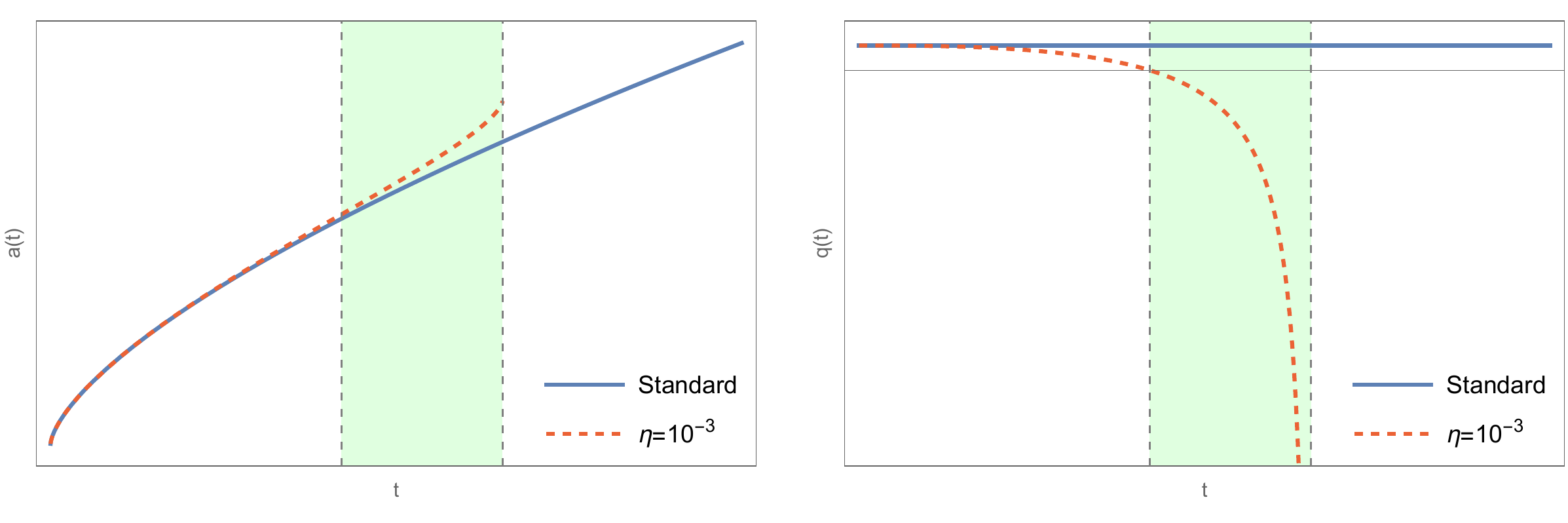}
	\caption{In left plot, the scale factor as a function of time for a choice of $w=0$. In right plot, the deccelaration parameter $q$ as a function of time for a choice of $w=0$.
 For both cases, in blue solid line, we show the standard case, and in red dashed line the result obtained after solving \eqref{frieda} and \eqref{qeq} with a value of $\eta=10^{-3}$. The green area corresponds to the zone when acceleration happens, according to the condition \eqref{conditionsacc}.}
	\label{Fig2}
\end{figure}

We solve numerically \eqref{frieda} considering $\rho_0=1$ just to get the functional form of $a$, and with the choice of $\eta=10^{-3}$ as example, we get the plots for $a$ and $q$ displayed in the left and right plots in Fig.~\ref{Fig2} respectively, where we compare with the standard result.  

We observe in both figures that in this model it is possible to have an acceleration expansion at late time. Notice that the result at the end of the accelerated expansion the scale factor becomes imaginary, something similar to the results found in \cite{Reyes:2022dil} for a choice of $s^{\mu\nu}$. This behaviour was expected from the denominator in \eqref{frieda} because to solve it is required to take the square root, and it is only real if $1-2a^4\eta>0$. 
\section{Spontaneous background fields}\label{SectionVI}
The consideration of spontaneous backgrounds in cosmology date back from the 
late eighties~\cite{Kostelecky:1989jp,Kostelecky:1989jw}. In that occasion, it was used to study a model in cosmology
inspired from string theory with two expansion scales. 
In the following, we will be interested to study a bumblebee model 
in the context of cosmology using Friedmann spacetimes.
\subsection{The bumblebee model}\label{SeccionVI:SubsectionA}
Consider the action
\begin{align} 
    S_B&=S_{\mathrm{EH}}+S_B+S_m  \,,
\end{align}
with
\begin{align} \label{Eq:Bumblebee_action}
    S_B&=\int_{\mathcal{M}}\mathrm{d}^4x \sqrt{-g} \bigg[ \frac {1}{ 2\kappa} \big(\prescript{(4)}{}{R}+ \xi   
    B^\mu B^\nu     \prescript{(4)}{}{R}_{\mu\nu} \big)-\frac{1}{4}B^{\mu\nu}B_{\mu\nu} -V( B_{\mu}B^{\mu} \pm b^2 )\bigg]  \,,
\end{align}
where $S_{\mathrm{EH}}$ is the Einstein-Hilbert action
and we have included a perfect fluid source $S_m$.
Also $B^{\mu}$ is the bumblebee field which in contrast to the nondynamical background has its 
own Euler-Lagrange equation of motion.
We define the field strength tensor of the bumblebee field
by
$B_{\mu\nu}=\partial_\mu B_\nu-\partial_\nu B_\mu$ and we are considering 
 $b^2=b_\mu b^\mu$ where $b_\mu$
is the VeV of the bumblebee field, that is to say, $\langle B_\mu \rangle =b_\mu$. Furthermore, 
$\xi$ is a coupling constant of the interaction term between the background and gravity.
We will take the potential to be
\begin{align}\label{Potential}
    V(B_{\mu}B^{\mu} \pm b^2)&=\frac{\lambda}{4} (B_{\mu}B^{\mu} \pm b^2 )  ^2\,,
\end{align}
where $\lambda$ is a coupling constant.
Alternatively, we may have obtained the action~\eqref{Eq:Bumblebee_action} 
by replacing
$u\to \frac14 \xi  B^{\mu}  B_{\mu}$,
$s^{\mu \nu}\to  \xi  \left(B^{\mu}  B^{\nu}  -  \frac14 g^{\mu \nu }  B^{\alpha}  B_{\alpha} \right)  $ and $t^{\mu \nu \rho \sigma}\to 0$ in
~\eqref{eq:explicit_diffeo}, together with installing kinetic and potential terms by hand.

We find the modified Einstein's equation
\begin{align} \label{Einstein-bumblebee}
    G^{\mu\nu}&=      (T_B)^{\mu\nu}+ \kappa (T_m)^{\mu\nu}  \,,
\end{align}
where $(T_m)^{\mu\nu}$ is the energy-momentum tensor of the perfect fluid~\eqref{E-M_perfectfluid} and 
\begin{align}\label{Energy-mom_bumblebee}
  (T_B)^{\mu\nu}&= \kappa\bigg[2V' B^\mu B^\nu+B^\mu_{\ \kappa}B^{\nu\kappa} -\bigg( V+\frac{1}{4}B^{\lambda\kappa}B_{\lambda\kappa} 
   \bigg)g^{\mu\nu}\bigg]+  \frac{\xi}{2} \bigg[B^\lambda B^\kappa R_{\lambda\kappa}g^{\mu\nu}
   \nonumber  \\
  &  -2\big(g^{\mu\rho}B^\nu +g^{\nu\rho}B^\mu\big) B^\sigma R_{\rho\sigma} + \nabla_\lambda\nabla^\mu 
  \big(B^\lambda B^\nu\big)+\nabla_\lambda\nabla^\nu \big(B^\lambda B^\mu\big) \nonumber \\
  & -\nabla_\kappa\nabla_\lambda \big(B^\lambda B^\kappa \big)g^{\mu\nu} -\nabla_\lambda\nabla^\lambda\big( B^\mu B^\nu\big)   \bigg] \,.
\end{align}

The equation of motion of the bumblebee field is
\begin{align} \label{bumblebee_field_eq}
    \nabla_\mu B^{\mu\nu}=2 V' B^\nu -\frac{\xi}{\kappa}B_\mu R^{\mu\nu}  \,,
\end{align}
where the prime denotes differentiation with respect to the argument of $V$. 

For the spontaneous breaking we also should include
a boundary term.
The boundary terms came
 from the first two terms in~\eqref{Eq:Bumblebee_action},
 which we decompose as follows. 
 We start by decomposing the field $B^\alpha$ into their normal and tangential projections
\begin{align}
   B^\alpha=e^\alpha_{a}B^a-n^\alpha B^\mathbf{n} \,.
\end{align}
Thus the decomposition for $B^\alpha B^\beta$ becomes
\begin{align}
  B^\alpha B^\beta = e^\alpha_{b}e^\beta_{b}B^aB^b -e^\alpha_{a}n^\beta B^a B^\mathbf{n}-
    n^\alpha e^\beta_{b}B^\mathbf{n} B^b+n^\alpha n^\beta \big(B^\mathbf{n}\big)^2  \,,
\end{align}
and the decomposition of the Ricci tensor can be shown to be
\begin{eqnarray}
 R_{\beta\delta}&=& \widetilde{E}_\beta^{ b}\widetilde{E}_\delta^{ d}\Big(R_{bd}-2K_{be}K^e_{\ d}+KK_{bd}
     -D_d a_b-a_d a_b + \frac{1}{N} e^{\rho}_{b}e^{\sigma}_{d}\mathcal{L}_m (\widetilde{E}_{\rho}^{e}\widetilde{E}_{\sigma}^{f}K_{ef})\Big) \nonumber \\
     & &  -\widetilde{E}_\beta^{ b}n_\delta \Big(D^e K_{eb}- D_b K\Big)- n_\beta \widetilde{E}_\delta^{d}\Big(D^e K_{ed} - D_d K\Big)\nonumber \\
     && +n_\beta n_\delta\Big(D_e a^e+a_e a^e+ K^{ ef}K_{ef}  - \frac{1}{N} q^{bd}e^{\rho}_{b}e^{\sigma}_{d}\mathcal{L}_m (\widetilde{E}_{\rho}^{e}\widetilde{E}_{\sigma}^{f}K_{ef})\big)\Big)\,.
\end{eqnarray}
 With these elements, we have 
\begin{eqnarray}
  R_{\beta\delta}B^\beta B^\delta &=&B^aB^b\Big(R_{ab}-2K_{ac}K^c_{\ b}+KK_{ab}-D_b a_a-a_b a_a 
    + \frac{1}{N} e^{\rho}_{a}e^{\sigma}_{b}\mathcal{L}_m (\widetilde{E}_{\rho}^{e}\widetilde{E}_{\sigma}^{f}K_{ef})\Big) \nonumber \\
     & & -B^a B^\mathbf{n} \Big(2\big(D^c K_{ca}- D_a K\big)\Big) +\big(B^\mathbf{n}\big)^2\Big(D_a a^a+a_a a^a+ K^{ ab}K_{ab} \notag \\&&
     - \frac{1}{N} q^{ab}e^{\rho}_{a}e^{\sigma}_{b}\mathcal{L}_m (\widetilde{E}_{\rho}^{e}\widetilde{E}_{\sigma}^{f}K_{ef})\Big) \,,
\end{eqnarray}
 and recalling Eq.~\eqref{EH_decom} leads to the standard boundary term which
 we include in the final expression.
 
By using the identity
\begin{align}
 e^{\rho}_{a}e^{\sigma}_{b}  & \Big[B^aB^b-\big(B^\mathbf{n}\big)^2q^{ab}\Big]\frac{1}{N} \mathcal{L}_m (\widetilde{E}_{\rho}^{e}\widetilde{E}_{\sigma}^{f}K_{ef}) = \nabla_\lambda \bigg[ n^\lambda\Big(B^aB^b -\big(B^\mathbf{n}\big)^2q^{ab}\Big)  K_{ab}\bigg] \notag
  \\&-\Big[B^aB^b-\big(B^\mathbf{n}\big)^2q^{ab}\Big]KK_{ab}  -\frac{1}{N}\widetilde{E}_{\rho}^{e}\widetilde{E}_{\sigma}^{f}K_{ef}\mathcal{L}_m\bigg[e^{\rho}_{a}e^{\sigma}_{ b}
   \Big(B^aB^b-\big(B^\mathbf{n}\big)^2q^{ab}\Big) \bigg] \,,
\end{align}
and the previous relations, we find the extended boundary term 
\begin{equation}
S_{\substack{\mathrm{ext} \\ \mathrm{GHY}}}=\frac{1}{2\kappa} \oint_{\partial\mathcal{M}}   \mathrm{d} ^3 y  \, \varepsilon
    \left(2K+\xi\Big(B^aB^b-\big(B^\mathbf{n}\big)^2h^{ab}\Big)  K_{ab}  \right)  \,.
\end{equation}
Notice that in the case when the interaction term vanishes we obtain the usual GHY boundary term.
\subsection{Modified Friedmann equations for $B_{\mu}=(0,B_{a} (x))$} \label{SectionVII}
Here we decompose the modified Einstein's equation,
by using 
our strategy in order to find the modified Friedmann equations.

We introduce a simplification by considering a purely tangential bumblebee field
\begin{align}
   B^\alpha=e^\alpha_{a}B^a \,,
\end{align}
and obtain the projection for the bumblebee field strength 
\begin{eqnarray}
B_{\mu\nu} =\widetilde{E}_\mu^{a}\widetilde{E}_\nu^{b}B_{ab}-n_\mu \widetilde{E}_\nu^{b}
\big[\Xi_b\big] +\widetilde{E}_\mu^{a}n_\nu\big[\Xi_a] \,,
\end{eqnarray}
where
\begin{eqnarray}
  B_{ab}=D_a B_b -D_b B_a\,,
\end{eqnarray}
and
\begin{eqnarray}
   \Xi_c=\frac{1}{N}e^\kappa_{c} \mathcal{L}_m\big(\widetilde{E}_\kappa^{d}B_d\big)\,.
\end{eqnarray}
By using the decomposition for the second covariant derivative of the product of bumblebee fields, see the 
Appendix~\eqref{AppendixD}, we can decompose the bumblebee energy-momentum tensor in their normal and tangential projections
as
\begin{eqnarray}
 (T_B)^{\mu\nu}= e^\mu_{a}e^\nu_{b}(T^B_1)^{ab} -(e^\mu_{a}n^\nu+n^\mu e^\nu_{a})(T^B_2)^a+  n^\mu n^\nu (T^B_3) \,,
\end{eqnarray}
with 
\begin{align}
   (T_1^B)^{ab}&=\kappa\bigg[  2V'B^aB^b+B^{ac}B^{b}_{\ c}-q^{ac}q^{bd}\mathcal{L}_tB_c\mathcal{L}_tB_d   -\bigg( V+\frac{1}{4}\big(B_{cd}B^{cd}-  2h^{cd}\mathcal{L}_tB_c\mathcal{L}_tB_d\big)  \bigg)q^{ab}\bigg] \nonumber \\
      & +\frac{\xi}{2}\bigg[q^{ab}\bigg(\frac{2k}{a(t)^2}B^c B_c -D_cD_d(B^cB^d)  -H\mathcal{L}_t (B_cB^c)\bigg) +D_cD^a(B^cB^b)+D_cD^b(B^cB^a)   \nonumber \\
      & -D^2(B^aB^b)+ q^{bc}\bigg(   \mathcal{L}_t  \mathcal{L}_t (B^aB_c)+3H\mathcal{L}_t (B^aB_c)   -2   \bigg(\frac{4k}{a(t)^2} +(\dot{H}+3H^2)\bigg)B^aB_c\bigg]\,,
\end{align}
\begin{align} \label{momentum_constraint_bumblebee}
   (T_2^B)^a   &= \kappa\bigg[B^{ac}\mathcal{L}_tB_c\bigg]+\frac{\xi}{2}\bigg[D_c\big(\mathcal{L}_t (B^cB^a)\big)-HD^a(B^cB_c)+2HD_c(B^cB^a)\bigg] \,,
\end{align}
\begin{align}
   (T_3^B)&=\kappa\bigg[ V+\frac{1}{4}\bigg(B_{cd}B^{cd}+  2q^{cd}\mathcal{L}_tB_c\mathcal{L}_tB_d\big)\bigg)  \bigg]
    -\frac{\xi}{2}\bigg[\frac{2k}{a(t)^2}B^c B_c  +H\mathcal{L}_t (B^cB_c)\nonumber \\
 & -D_cD_d(B^cB^d)\bigg] \,.
\end{align}
Taking into consideration $(T_m)^{\mu\nu}$ and $G^{\mu\nu}$ for the perfect fluid and the
 FLRW metric, we obtain the modified Einstein's equation 
\begin{align}
  G^{ab}&=\kappa\bigg[  2V'B^aB^b+B^{ac}B^{b}_{\ c}-q^{ac}q^{bd}\mathcal{L}_tB_c\mathcal{L}_tB_d   -\bigg( V+\frac{1}{4}\big(B_{cd}B^{cd}-  2q^{cd}\mathcal{L}_tB_c\mathcal{L}_tB_d\big)  \bigg)q^{ab}\bigg] \nonumber \\
      &+\frac{\xi}{2}\bigg[q^{ab}\bigg(\frac{2k}{a(t)^2}B^c B_c -D_cD_d(B^cB^d) -H\mathcal{L}_t (B_cB^c)\bigg) +D_cD^a(B^cB^b)+D_cD^b(B^cB^a) \nonumber \\
      &-D^2(B^aB^b)+ \mathcal{L}_t^2 (B^aB^b)+7H\mathcal{L}_t (B^aB^b) -4   \bigg(\frac{2k}{a(t)^2} -H^2\bigg)B^aB^b\bigg]+\kappa(T_m)^{ab}\,,
\end{align}
and 
\begin{align}
G^{\mathbf{n}\mathbf{n}} &=\kappa\bigg[ V+\frac{1}{4}\bigg(B_{cd}B^{cd}+ 
 2q^{cd}\mathcal{L}_tB_c\mathcal{L}_tB_d\big)\bigg)  \bigg] -\frac{\xi}{2}
 \bigg[\frac{2k}{a(t)^2}B^c B_c  +H\mathcal{L}_t (B^cB_c)-D_cD_d(B^cB^d)\bigg] \nonumber \\
 &+\kappa(T_m)^{\mathbf{n}\mathbf{n}}\,.
\end{align}
Also, the bumblebee field equations \eqref{bumblebee_field_eq} 
can be decomposed in their spatial and normal projections as 
\begin{align} \label{bumblebee_field_eq_projected}
      & e^\nu_{b}\Bigg[D_a (B^{ab})-Hq^{ba}\mathcal{L}_tB_a-q^{ba} \mathcal{L}_t^2B_a -2 \bigg[V' -\frac{\xi}{2\kappa}\bigg(\frac{2k}{a(t)^2} + (\dot{H}+3H^2)\bigg)\bigg]B^b\Bigg]  \nonumber \\
       &-n^\nu  \Big[-  D^c \big(\mathcal{L}_tB_c\big)\Big]=0\,.
\end{align}
At this point we make another simplification. By looking the normal projection 
of the bumblebee field equation we take as ansatz the condition $D_m B_n=0$ 
meaning that the bumblebee field will only depends on time $B_a=B_a(t)$. 

Under this consideration we obtain the first modified Friedmann equations 
\begin{align}
H^2 &= \frac{1}{3\Big(1-\frac{\xi}{3}B^cB_c\Big)}\bigg(\kappa \rho+\kappa\bigg[ V+\frac{1}{2}  q^{cd}\dot{B}_c\dot{B}_d 
 \bigg]-\frac{\xi}{2}\bigg[\frac{2k}{a(t)^2}B^c B_c  +2HB^c\dot{B}_c\bigg] \\ &  -\frac{3k}{a(t)^2}\bigg) \notag \,,
\end{align}
the second Friedmann equation
\begin{align}
  \dot{H}+H^2&= \frac{1}{(-6)\Big(  1  -\frac{\xi}{3}B^cB_c\Big)}\bigg(\kappa (3P+\rho)+\kappa\Big[   2V'B^cB_c-2V+q^{cd}\dot{B}_c\dot{B}_d  \Big] \nonumber \\
       &+\frac{\xi}{2}\bigg[-\frac{4k}{a(t)^2}B^c B_c+4H^2B^c B_c -10Hq^{cd}\dot{B}_cB_d+2q^{cd}\ddot{B}_cB_d+2q^{cd}\dot{B}_c\dot{B}_d\bigg]\bigg) \,,
\end{align}
 and the condition extra
\begin{align}
    (T^B)^{\langle ab \rangle}&=\kappa\bigg[  2V'B^aB^b-q^{ac}q^{bd}\dot{B}_c\dot{B}_d-\frac{1}{3}q^{ab}\bigg(  2V'B^cB_c-q^{cd}\dot{B}_c\dot{B}_d\bigg)\bigg] \,,\nonumber \\
      &+\frac{\xi}{2}\bigg[ q^{bc}\bigg(\mathcal{L}_t^2(B^aB_c)+3H\mathcal{L}_t (B^aB_c)-2   \bigg(\frac{4k}{a(t)^2} +(\dot{H}+3H^2)\bigg)B^aB_c\bigg)\nonumber \\
      &-\frac{1}{3}q^{ab}\bigg(\mathcal{L}_t^2 (B^cB_c)+3H\mathcal{L}_t (B^cB_c)-2   \bigg(\frac{4k}{a(t)^2}+(\dot{H}+3H^2)\bigg)B^cB_c\bigg)\bigg]=0\,.
    \end{align}
Together with the bumblebee field equations
\begin{align}
  \ddot{B}_a+ H\dot{B}_a+2 \bigg[V' -\frac{\xi}{2\kappa}\bigg(\frac{2k}{a(t)^2} + (\dot{H}+3H^2)\bigg)\bigg]B_a=0  \,.
\end{align}
We provide the details of the decomposition in the Appendix~\eqref{AppendixD}.
\subsubsection{The case $\xi=0$}
We turn off the interaction term with $\xi=0$ and
the bumblebee field equation of motion simplifies to
\begin{eqnarray}
  \ddot{B}_a+H\dot{B}_a+2V' B_a =0 \,.
\end{eqnarray}
We have the first modified Friedmann equation
\begin{eqnarray}
H^2 &=&\frac{1}{3}\bigg\lbrace \kappa\rho+\kappa\bigg[V+\frac{1}{2}q^{ab}\dot{B}_a \dot{B}_b \bigg] -\frac{3k}{a(t)^2}\bigg\rbrace \,,
\end{eqnarray}
and the second modified Friedmann equation
\begin{eqnarray}
 \dot{H}+H^2&=&\frac{1}{(-6)}\bigg\lbrace \kappa(3P+\rho)+\kappa\bigg[2V' q^{ab}B_aB_b-2V +q^{ab}\dot{B}_a \dot{B}_b\bigg] \bigg\rbrace \,,
\end{eqnarray}
and the homogeneity-isotropy condition
\begin{eqnarray}
  0&=& \kappa\bigg[\big(2V'B_aB_b-\dot{B}_a\dot{B}_b\big)  -\frac{1}{3} q_{ab}q^{cd}\big(2V'B_cB_d-\dot{B}_c\dot{B}_d\big)\bigg]  \,.
\end{eqnarray}
For $V'> 0$ the isotropy-homogeneity condition has the general solution 
\begin{eqnarray} \label{bumblebee-ansatz}
    \dot{B}_a=\pm \sqrt{2V'}B_a\,.
\end{eqnarray}
By taking the time derivative, we obtain an expression for $\ddot{B}_a$ as follows 
\begin{eqnarray}
    \ddot{B}_a&=&\pm \sqrt{2}\bigg(\frac{1}{2}\frac{1}{\sqrt{V'}}\frac{\mathrm{d}V'}{\mathrm{d}(B^c B_c)}\frac{\partial (B^bB_b)}{\partial t}B_a+\sqrt{V'}\dot{B}_a\bigg)    \nonumber \\
    &=&\pm \sqrt{2}\bigg(\frac{1}{2}\frac{V''}{\sqrt{V'}}\frac{\partial (B^bB_b)}{\partial t}B_a+\sqrt{V'}\dot{B}_a\bigg)  \\
    &=&\pm 2\bigg(\frac{V''}{\sqrt{2V'}}\big(-H\pm\sqrt{2V'}\big)h^{bc}B_bB_c\pm V'\bigg)B_a\notag \,.
\end{eqnarray}
Using this expression in the bumblebee field equation we obtain
\begin{align}
 \bigg(2V''\big(\mp H+(2V')^{1/2}\big)q^{bc}B_bB_c\pm H(2V')+2(2V')^{3/2}\bigg)B_a=0 \,.
\end{align}
Thus the final equation to solve becomes an scalar equation
\begin{align}
   2V''\big(\mp H+(2V')^{1/2}\big)B^2\pm H(2V')+2(2V')^{3/2}=0\,.
\end{align}
Lets recall the quartic potential 
\begin{align}
    V(B^2)&=\frac{\lambda}{4} (B^2- b^2)^2,\\ B^2&=B^aB_a \,.
\end{align}
We obtain
\begin{eqnarray}
      \mp H \lambda b^2+( \lambda (B^2-b^2))^{1/2}( \lambda (3B^2-2b^2))=0 \,,
  \end{eqnarray}   
which lead to  
 \begin{eqnarray}    
     (B^2-b^2)^{1/2}(3B^2-2 b^2)=\pm \frac{H}{\sqrt{\lambda}}b^2\,.
\end{eqnarray}
Defining the quantity $A^2=B^2-b^2$ we obtain
\begin{eqnarray}
    A^2\vert A\vert+\frac{b^2}{3}\vert A\vert\mp \frac{H}{3\sqrt{ \lambda }}b^2=0 \,.
\end{eqnarray}
We recognize this equation as the depressed cubic equation
\begin{eqnarray}
    A^3+pA+q=0\,,
\end{eqnarray}
with
\begin{eqnarray}
    p&=&\frac{b^2}{3}\,, \\
    q&=&\mp\frac{H}{3\sqrt{\lambda}}b^2 \,.
\end{eqnarray}
The equation os well know for having two types of solutions: for three real solutions we have the Viette formula, and for one real solution and two complex solutions we have the Cardano formula. In this case the discriminant becomes
\begin{eqnarray}
    \Delta=-(4p^3+27q^2)=-b^4\bigg(\frac{4}{27}b^2+3\frac{H^2}{\lambda}\bigg)< 0\,.
\end{eqnarray}
Thus we have one real root and two complex roots for the cubic polynomial. The Cardano formula gives the real solution
\begin{eqnarray}
    A&=&\bigg( -\frac{q}{2}+\sqrt{\frac{q^2}{4}+\frac{p^3}{27}}  \bigg)^{1/3} +\bigg( -\frac{q}{2}-\sqrt{\frac{q^2}{4}+\frac{p^3}{27}} \bigg)^{1/3}\nonumber \\
    &=&\bigg(\pm \frac{H}{6\sqrt{\lambda}}b^2+\frac{b^2}{3}\sqrt{ \frac{H^2}{4\lambda}+\frac{b^2}{81}}  \bigg)^{1/3} +\bigg(\pm \frac{H}{6\sqrt{ \lambda }}b^2-\frac{b^2}{3}\sqrt{ \frac{H^2}{4 \lambda }+\frac{b^2}{81}} \bigg)^{1/3}\,.
\end{eqnarray}
Obtaining the time dependent bumblebee field solution
\begin{eqnarray}
  B^2(t)&=&\bigg(\frac{b^2}{6\sqrt{ \lambda }}\bigg)^{2/3}\Bigg[\bigg(\pm H+\sqrt{ H^2+\frac{4 \lambda b^2}{81}}  \bigg)^{2/3} +\bigg(\pm H-\sqrt{ H^2+\frac{4 \lambda b^2 }{81}}  \bigg)^{2/3}\Bigg]-\frac{7}{9}b^2 \,.\notag \\
\end{eqnarray}
Therefore, the first modified Friedmann equations can be written as
\begin{eqnarray}
 H^2  &=& \frac{1}{3}\bigg(\kappa \rho+\frac{\lambda\kappa}{4}( 3B^dB_d-b^2  )(B^cB_c-b^2)-\frac{3k}{a(t)^2}\bigg) \,,
\end{eqnarray}
and the second
\begin{eqnarray}
   \dot{H}+H^2&=&  \frac{1}{(-6)}\bigg[\kappa (3P+\rho)+\frac{\lambda\kappa}{2}(3B^dB_d+ b^2)(B^cB_c-b^2)\bigg] \,.
\end{eqnarray}
Solving the nonlinear equations is beyond the scope of the present work, and we leave such analysis for a
future study.
\subsubsection{Nonminimal coupling extended case $\xi\neq 0$}
Inspired by the advances made in stating the differential equations 
in the previous part we extend to consider $\xi\neq 0$. 

Lets consider the tangential bumblebee field equation
\begin{eqnarray}
     \ddot{B}_a=-\bigg(H\dot{B}_a+2 \bigg[V' -\frac{\xi}{2\kappa}\bigg(\frac{2k}{a(t)^2} + (\dot{H}+3H^2)\bigg)\bigg]B_a  \bigg) \,.
\end{eqnarray}
Replacing in the homogeneity-isotropy equation we obtain
 \begin{align}
        &\kappa\bigg[  2V'B_aB_b-\dot{B}_a\dot{B}_b-\frac{1}{3}q_{ab}\bigg(  2V'B^cB_c-q^{cd}\dot{B}_c\dot{B}_d\bigg)\bigg]\nonumber \\
      &+\frac{\xi}{2}\bigg[2\dot{B}_a\dot{B}_b-2H\dot{B}_aB_b-2HB_a\dot{B}_b -4 \bigg[V'-\frac{\xi}{2\kappa}\bigg(\frac{2k}{a(t)^2} \nonumber \\
      &+ (\dot{H}+3H^2)\bigg)+\bigg(\frac{2k}{a(t)^2} +(\dot{H}+2H^2)\bigg)\bigg]B_a  B_b \nonumber \\
      &-\frac{1}{3}q_{ab}\bigg(2q^{cd}\dot{B}_c\dot{B}_d -4Hq^{cd}\dot{B}_cB_d  -4   \bigg[V' -\frac{\xi}{2\kappa}\bigg(\frac{2k}{a(t)^2} \nonumber \\
      &+ (\dot{H}+3H^2)\bigg)+\bigg(\frac{2k}{a(t)^2} +(\dot{H}+2H^2)\bigg)\bigg]B^cB_c\bigg)\bigg]=0 \,.
    \end{align}
Obtaining a tensorial equation with lower degrees in time derivatives. We consider the tangential 
bumblebee field equation for $\xi=0$ \eqref{bumblebee-ansatz} as an ansatz at this point. Hence,
we obtain a major simplification for the isotropy-homogeneity equation
\begin{align}
   & -2\xi\bigg[H\sqrt{2V'}-\frac{\xi}{2\kappa}H^2+\bigg(\frac{2k}{a(t)^2} +(\dot{H}+2H^2)\bigg)\bigg(1-\frac{\xi}{2\kappa}\bigg)\bigg]\nonumber \\
   & \bigg(B_aB_b-\frac{1}{3}q_{ab}B^cB_c\bigg)=0 \,.
\end{align}
Notice that this case is disconnected from the $\xi=0$ case. The solution for the second parenthesis
 correspond to a null bumblebee field. The only possibility is to solve the scalar equation
 \begin{align}
       H\sqrt{2V'} + \bigg(1-\frac{\xi}{2\kappa}\bigg)\bigg(\frac{2k}{a(t)^2} + (\dot{H}+2H^2)\bigg)-\frac{\xi}{2\kappa}H^2=0 \,.
    \end{align}
Considering again the quartic potential and solving the term with the derivative of the potential we obtain a new equation 
for the bumblebee field
\begin{align}
    B^2&=\frac{1}{H^2}\Bigg[\bigg(1-\frac{\xi}{2\kappa}\bigg)\bigg(\frac{2k}{a(t)^2} + (\dot{H}+2H^2)\bigg)-\frac{\xi}{2\kappa}H^2\Bigg]^2+ b^2\,,
\end{align}
and a new set of modified Friedmann equations
    \begin{align}
 H^2  &= \frac{1}{3\Big(1-\frac{\xi}{3}\big(\frac{1}{H^2}\big[(\dot{H}+2H^2)-\frac{\xi}{2\kappa}(\dot{H}+3H^2)\big]^2\mp b^2\big)\Big)}\\
 &
 \bigg(\kappa \rho +\frac{\lambda\kappa}{4H^2}\bigg[(\dot{H}+2H^2)-\frac{\xi}{2\kappa}(\dot{H}+3H^2)\bigg]^2\bigg[ \frac{3}{H^2}\bigg[(\dot{H}+2H^2)-\frac{\xi}{2\kappa}(\dot{H}+3H^2)\bigg]^2\mp 2b^2  \bigg]\nonumber \\
 &+\xi\bigg[(\dot{H}+2H^2)-\frac{\xi}{2\kappa}(\dot{H}+3H^2)\bigg]\bigg(\frac{1}
 {H^2}\bigg[(\dot{H}+2H^2)-\frac{\xi}{2\kappa}(\dot{H}+3H^2)\bigg]^2\mp b^2\bigg)\bigg)\notag  \,,
 \end{align}
 and
 \begin{align}
   \dot{H}+H^2&=  \frac{1}{(-6)\Big(  1  -\frac{\xi}{3}\big(\frac{1}{H^2}\big[(\dot{H}+2H^2)-\frac{\xi}{2\kappa}(\dot{H}+3H^2)\big]^2\mp b^2\big)\Big)}
 \\
   &  \bigg(\kappa (3P+\rho)+\frac{\lambda\kappa}{2H^2}\bigg[(\dot{H}+2H^2)-\frac{\xi}{2\kappa}(\dot{H}+3H^2)\bigg]^2   \bigg[   \frac{3}{H^2}\bigg[(\dot{H}+2H^2)\notag \\
   &-\frac{\xi}{2\kappa}(\dot{H}+3H^2)\bigg]^2\mp 4b^2  \bigg]+2\xi\bigg[(3\dot{H}+7H^2) -\frac{\xi}{\kappa}(\dot{H}+3H^2)\bigg]
   \nonumber \\
      &  \bigg(\frac{1}{H^2}\bigg[(\dot{H}+2H^2)-\frac{\xi}{2\kappa}(\dot{H}+3H^2)\bigg]^2\mp b^2\bigg)\bigg)\notag  \,.
\end{align}
Again we leave for future work the finding of solutions of the above set of equations.
\section{Conclusions}\label{SectionVIII}
We have studied explicit and spontaneous background fields, highlighting the challenges 
associated with explicit symmetry breaking. We have shown why these issues do 
not arise with dynamical backgrounds, as they are governed by their equations of motion.
However, there are cases for explicit breaking in which the problem can be avoided 
for example restricting the geometry as in the case of Chern-Simons extensions and some sectors of the minimal gravitational SME. 
We have focused on cosmology in the presence of both explicit and dynamical background fields. We have demonstrated that in cosmological models relying on isotropic and homogeneous spacetime, specifically for the $t$-sector of the SME
there are no issues with the no-go result.
 By employing the formalism of maximally symmetric subspaces, we have derived additional conditions that must be imposed on the energy-momentum tensor of the background fields. These conditions establish a set of stringent criteria for the backgrounds to maintain form invariance under isotropic and homogeneous transformations.
 
 In the $t$ sector of the gravitational minimal Standard Model Extension (SME) and the dynamical bumblebee model, we have identified boundary terms that depend on the backgrounds coupled to the extrinsic tensor. These terms enable a proper Dirichlet formulation within the variational formalism.
We also have derived the modified Friedmann equations in both models. These are based on the $3+1$ decomposition of spacetime 
and a technique to perform the decomposition of the modified Einstein's equation.
For the $t$ sector we have found possible configurations of the background field that allow to accelerate the universe in the absence of standard dark energy through the cosmological constant. We have found the dynamical equations in the bumblebee model, leaving their analysis due to their non linear character to be considered in another work.

In addition to other approaches for addressing explicit breaking of diffeomorphism, we have shown 
that implementing a non-dynamical background field can be achieved consistently by imposing symmetry restrictions in a symmetric  subspace.
\medskip

\acknowledgments
We acknowledge J. Belinchon for insightful discussions on Noether, Killing, and 
Homotheties fields. Our thanks also go to A. Diez-Tejedor for valuable comments on the 
bumblebee model and for directing us to complementary references. We are especially grateful 
to A. Kostelecky and R. Lehnert for their kind hospitality during the third IUCSS Workshop on 
Theoretical Aspects of the Standard-Model Extension and for their significant questions on 
isotropy and homogeneity. CMR appreciates partial support from the Fondecyt Regular research project 1241369. 
CR acknowledges support from the ANID fellowship No. 21211384 and Universidad de Concepción.
\appendix

\section{Basic identities}\label{AppendixA}
In this appendix, we present the basic geometric objects
 that we use to decompose the equations of motion. 
We introduce a method that serves in both models, explicit and spontaneous breaking.

Consider the expression~\eqref{delta_strat}, we have for a vector
\begin{align}
    V^{\alpha}&=e^{\alpha}_a V^a-n^{\alpha} V^{\bf n}  \,,
\end{align}
where
\begin{align}
   V^a&=\widetilde{E}^a_{\mu} V^{\mu} \,, \\  V^{\bf n}&=n_{\mu} V^{\mu} \,,
\end{align}
and a $1$-form as
\begin{align}
     \omega_{\alpha}&=\widetilde{E}_{\alpha}^a \omega_a-n_{\alpha} \omega_{\bf n}    \,,
\end{align}
with
\begin{align}
   \omega_a&= e_a^{\mu} \omega_{\mu}\,, \\
     \omega_{\bf n}&=n^{\mu} \omega_{\mu}   \,.
\end{align}
By definition for a tensor living
 in the hypersurface  $\omega_{\alpha}=\widetilde{E}^a_{\alpha}\omega_a$~\cite{Poisson:2004}, we have
\begin{align}
  D_b \omega_a=e^\alpha_{a}e^\beta_{b}\nabla_\beta \omega_\alpha \,.
\end{align}
For a tangential vector $T^\alpha=e^\alpha_a T^a$ one has
\begin{eqnarray}
   D_b T^a = \widetilde{E}_\alpha^{a}e^\beta_{b}\nabla_\beta T^\alpha  \,,
\end{eqnarray}
 which can be proved as follows. From
\begin{eqnarray}
\widetilde{E}_\alpha^{a}e^\beta_{b}\nabla_\beta T^\alpha&=&e^\beta_{b}\nabla_\beta (\widetilde{E}_\alpha^{a}T^\alpha)-e^\beta_{b}(\nabla_\beta \widetilde{E}_\alpha^{a})e^\alpha_{c}T^c \,.
\end{eqnarray}
we recognize the partial derivative on the hypersurface 
due to the chain rule while completing the covariant derivative as follows
\begin{eqnarray}
    \widetilde{E}_\alpha^{a}e^\beta_{b}\nabla_\beta T^\alpha&=&\partial_b T^a+e^\beta_{b}\widetilde{E}_\alpha^{a}(\nabla_\beta e^\alpha_{c})T^c \,. 
\end{eqnarray}
The tangential projection of the second term above 
is related to the Christoffel symbols for the induced metric, that is to say,
\begin{align}
  \Gamma^c_{\ ab}=\widetilde{E}_\alpha^{c}(\nabla_\beta e^{\alpha}_{a}) e^\beta_{b}\,,
\end{align}
which produces
\begin{align}
    \widetilde{E}_\alpha^{a}e^\beta_{b}\nabla_\beta T^\alpha&=D_bT^a \,.
\end{align}

The generalization for an arbitrary tangential tensor 
$T^{\alpha_1\alpha_2\dots\alpha_p}_{\beta_1\beta_2\dots\beta_q}$
\begin{align}\label{eq:indices_hypersurface}
   T^{\alpha_1\alpha_2\dots\alpha_p}_{\beta_1\beta_2\dots\beta_q}=
    e^{\alpha_1}_{ a_1}e^{\alpha_2}_{ a_2}\dots e^{\alpha_p}_{a_p}\widetilde{E}_{\beta_1}^{ b_1}
    \widetilde{E}_{\beta_2}^{ b_2}\dots \widetilde{E}_{\beta_q}^{ b_q}  \notag T^{a_1a_2\dots a_p}_{b_1b_2\dots b_q}  \,,
\end{align}
is given by the expression
\begin{align}  
 D_c T^{a_1a_2 \dots a_p}_{b_1b_2\dots b_q}=  \widetilde{E}_{\alpha_1}^{a_1} \widetilde{E}_{\alpha_2}^{a_2}
    \dots \widetilde{E}_{\alpha_p}^{a_p}e^{\beta_1}_{b_1}e^{\beta_2}_{b_2}\dots e^{\beta_q}_{b_q}
    e^\gamma_{c}    \nabla_\gamma    T^{\alpha_1\alpha_2\dots\alpha_p}_{\beta_1\beta_2\dots\beta_q}  \,.
\end{align}

Now, consider the Lie derivative of a tangential tensor along the $m^\mu=Nn^\mu$ direction
\begin{align}
 &  \mathcal{L}_m (e^{\mu_1}_{a_1}\dots e^{\mu_p}_{a_p}\widetilde{E}_{\nu_1}^{b_1}\dots \widetilde{E}_{\nu_q}^{b_q}T^{a_1\dots a_p}_{b_1\dots b_q})  
 =m^\lambda \nabla_\lambda (e^{\mu_1}_{a_1}\dots e^{\mu_p}_{a_p}\widetilde{E}_{\nu_1}^{b_1}
    \dots \widetilde{E}_{\nu_q}^{b_q}T^{a_1\dots a_p}_{b_1\dots b_q})    \\
    &  - \nabla_\lambda m^{\mu_1}(e^{\lambda}_{a_1}\dots e^{\mu_p}_{a_p}
    \widetilde{E}_{\nu_1}^{b_1}\dots \widetilde{E}_{\nu_q}^{b_q}T^{a_1\dots a_p}_{b_1\dots b_q})
    -\dots -\nabla_\lambda m^{\mu_p}(e^{\mu_1}_{a_1}\dots e^{\lambda}_{a_p}\widetilde{E}_{\nu_1}^{b_1}
    \dots \widetilde{E}_{\nu_q}^{b_q}T^{a_1\dots a_p}_{b_1\dots b_q})  \nonumber \\
    &+\nabla_{\nu_1}m^\lambda (e^{\mu_1}_{a_1}\dots e^{\mu_p}_{a_p}\widetilde{E}_{\lambda}^{b_1}
    \dots \widetilde{E}_{\nu_q}^{b_q}T^{a_1\dots a_p}_{b_1\dots b_q}) 
    +\dots+\nabla_{\nu_q}m^\lambda (e^{\mu_1}_{a_1}\dots e^{\mu_p}_{a_p}\widetilde{E}_{\nu_1}^{b_1}
    \dots \widetilde{E}_{\lambda}^{b_q}T^{a_1\dots a_p}_{b_1\dots b_q}) \notag \,,
\end{align}
and the basic elements
\begin{align}
    (\nabla_\lambda m^\mu)e^\lambda_{a}&= (m^\mu \nabla_\lambda \ln N + N\nabla_\lambda n^\mu )e^\lambda_{a} \nonumber \\
    &= (m^\mu D_a \ln N + N e^\lambda_{a} \nabla_\lambda n^\mu ) \nonumber \\
    &= N(n^\mu a_a +  e^\mu_{c}K^c_{\ a} ) \,,
\\
  (\nabla_\nu m^\lambda)\widetilde{E}_\lambda^{b}&=(m^\lambda \nabla_\nu \ln N+N\nabla_\nu n^\lambda) \widetilde{E}_\lambda^{b}\nonumber \\
   &=N\widetilde{E}_\lambda^{b} \nabla_\nu n^\lambda \nonumber \\
   &=N(\widetilde{E}_\nu^{c}K^b_{\ c}-n_\nu a^b) \,.
\end{align} 
where we have used the definition of the extrinsic curvature $K_{\mu \nu}=\nabla_{\mu} n_{\nu}+n_{\mu} a _{\nu}$ and 
introduced the acceleration four-vector $a_{\mu}= n^{\alpha}\nabla_{\alpha} n_{\mu}$.
Since the
 Lie derivative only act on Lorentzian indices included 
 in the projection of the tensor, the Lie derivate 
 can be written as
\begin{align}
&  \mathcal{L}_m (e^{\mu_1}_{a_1}\dots e^{\mu_p}_{a_p}\widetilde{E}_{\nu_1}^{b_1}
\dots \widetilde{E}_{\nu_q}^{b_q}T^{a_1\dots a_p}_{b_1\dots b_q})  
 =m^\lambda \nabla_\lambda (e^{\mu_1}_{a_1}\dots e^{\mu_p}_{a_p}\widetilde{E}_{\nu_1}^{b_1}\dots \widetilde{E}_{\nu_q}^{b_q}T^{a_1\dots a_p}_{b_1\dots b_q}) \\
    &\phantom{{}={}}\hspace{1.8cm} - N(n^{\mu_1}a_{a_1}+e^{\mu_1}_{c_1}K^{c_1}_{\ \ a_1})\dots e^{\mu_p}_{a_p}\widetilde{E}_{\nu_1}^{b_1}\dots \widetilde{E}_{\nu_q}^{b_q}T^{a_1\dots a_p}_{b_1\dots b_q}\nonumber \\
    &\phantom{{}={}}\hspace{1.8cm} -\dots -Ne^{\mu_1}_{a_1}\dots (n^{\mu_p}a_{a_p}+e^{\mu_p}_{c_p}K^{c_p}_{\ \ a_p})\widetilde{E}_{\nu_1}^{b_1}\dots \widetilde{E}_{\nu_q}^{b_q}
    T^{a_1\dots a_p}_{b_1\dots b_q}\nonumber \\
    &\phantom{{}={}}\hspace{1.8cm} +Ne^{\mu_1}_{a_1}\dots e^{\mu_p}_{a_p}(\widetilde{E}_{\nu_1}^{c_1}K^{b_1}_{\ \ c_1}-n_{\nu_1}a^{b_1})\dots 
    \widetilde{E}_{\nu_q}^{b_q}  T^{a_1\dots a_p}_{b_1\dots b_q} \nonumber \\
    &\phantom{{}={}}\hspace{1.8cm} +\dots+N e^{\mu_1}_{a_1}\dots e^{\mu_p}_{a_p}\widetilde{E}_{\nu_1}^{b_1}\dots 
     (\widetilde{E}_{\nu_q}^{c_q}K^{b_q}_{\ \ c_q}-n_{\nu_q}a^{b_q})T^{a_1\dots a_p}_{b_1\dots b_q} \,. \notag
\end{align}
Therefore, the projection on the hypersurface of the Lie derivative gives
\begin{align} \label{liederivative}
&\frac{1}{N}\widetilde{E}_{\mu_1}^{i_1}\dots \widetilde{E}_{\mu_p}^{i_p}\dots e^{\nu_1}_{j_1}\dots e^{\nu_q}_{j_q}  \mathcal{L}_m (e^{\mu_1}_{a_1}
\dots e^{\mu_p}_{a_p}\widetilde{E}_{\nu_1}^{b_1}\dots \widetilde{E}_{\nu_q}^{b_q}   T^{a_1\dots a_p}_{b_1\dots b_q}) \\
 &\phantom{{}={}}\hspace{1.8cm} =\widetilde{E}_{\mu_1}^{i_1}\dots \widetilde{E}_{\mu_p}^{i_p}\dots e^{\nu_1}_{j_1}\dots e^{\nu_q}_{j_q} n^\lambda\nabla_\lambda (e^{\mu_1}_{a_1}\dots e^{\mu_p}_{a_p}\widetilde{E}_{\nu_1}^{b_1}\dots \widetilde{E}_{\nu_q}^{b_q}
 T^{a_1\dots a_p}_{b_1\dots b_q}) \notag \\ &\phantom{{}={}}\hspace{1.8cm}  - K^{i_1}_{\ \ c_1}T^{c_1\dots i_p}_{j_1\dots j_q}-\dots -K^{i_p}_{\ \ c_p}T^{i_1\dots c_p}_{j_1\dots j_q}+K^{c_1}_{\ \ j_1}T^{i_1\dots i_p}_{c_1\dots j_q}+\dots+K^{c_q}_{\ \ j_q}T^{i_1\dots i_p}_{j_1\dots c_q} \notag \,. 
\end{align}
\section{Projections in the $t$ sector }  \label{AppendixB}
With the previous expressions, 
we are ready to compute the decomposition of the derivatives acting on  $t^{\mu\nu\rho\sigma}$.
We use the following strategy
to decompose an arbitrary tensor $G^{\mu_1 \mu_2\dots \mu_n}$: 
\begin{itemize}
    \item As the first step we decompose the tensor  $G^{\mu_1 \mu_2\dots \mu_n}$ in its tangential and normal projections. 
    \item We apply the covariant derivative on the decomposed tensor focusing on the derivative terms of tangential components. Additionally, terms like $\nabla_\mu n^\nu$ can be decomposed into the extrinsic curvature $K_{ab}$ and the acceleration vector $a^a$.
    \item In cases where the free indices of an expression are inside a covariant derivative, meaning they aren't factored out as projectors outside the expression, we will introduce Kronecker deltas and the completeness relation to extract these indices outside of the covariant derivative.
    \item In terms like $n_\mu \nabla_\nu \big(e^\mu_a \dots\big)$ we complete the derivative using the orthogonality of the projectors, allowing us to rewrite it as
  $e^\mu_a \nabla_\nu n_\mu \dots $. This form can be identified with geometric quantities and has the projectors with free indices on the outside.  
    \item Given that we have previously obtained the normal and tangential projections of the covariant derivative of tangential tensors, we use these expressions to identify Lie derivatives and covariant derivatives on the hypersurface. This allows us to derive an expression that contains well-defined quantities on the hypersurface, accompanied by the corresponding projectors for each free index in their normal and tangential directions.
\end{itemize}

We can show the steps by computing the covariant derivative of a contravariant vector. Lets focus on
\begin{align}
    G^\alpha=e^\alpha_{a}G^a-n^\alpha G^\mathbf{n} \,.
\end{align}
Taking the covariant derivative of $G^\alpha$ we obtain
\begin{align}
   \nabla_\beta G^\alpha&=\nabla_\beta\big(e^\alpha_{ a}G^a\big)-(\nabla_\beta n^\alpha)
    G^\mathbf{n}-n^\alpha \nabla_\beta G^\mathbf{n}\,,
\end{align} 
and using the definition of the extrinsic curvature
\begin{align}
    \nabla_\mu n_\nu&= \widetilde{E}_\mu^a \widetilde{E}_\nu^b K_{ab}- n_\mu a_\nu \,,
\end{align}
we have    
\begin{align}
    \nabla_\beta G^\alpha&=\nabla_\beta\big(e^\alpha_{ a}G^a\big)-\left(e^\alpha_{ a}K^a_{\ b}  \widetilde E_\beta^{ b}
    -n_\beta a^\alpha \right) G^\mathbf{n}-n^\alpha \nabla_\beta G^\mathbf{n} \,.
  \end{align} 
Now we introduce two Kronecker deltas in the first term due to two unprojected indices
 and in the last term we introduce one delta. As described we have as many deltas as unprojected indices has the term,
 yielding
  \begin{align} 
   \nabla_\beta G^\alpha&= \delta^\alpha_{\ \gamma}\delta^\delta_{\ \beta}
  \nabla_\delta\big(e^\gamma_{a}G^a\big)-(e^\alpha_{a}K^a_{\ b}\widetilde{E}_\beta^{b} 
  -n_\beta a^\alpha) G^\mathbf{n}-n^\alpha \delta^\delta_{\ \beta}\nabla_\delta G^\mathbf{n} \,,
 \end{align} 
and by using the completeness relation we obtain
  \begin{align} 
   \nabla_\beta G^\alpha&= (e^\alpha_{c}\widetilde{E}_\gamma^{c}-n^\alpha n_\gamma)
   (e^\delta_{d}\widetilde{E}_\beta^{d}-n^\delta n_\beta)\nabla_\delta\big(e^\gamma_{a}V^a\big)
   -(e^\alpha_{a}K^a_{\ b}\widetilde{E}_\beta^{b}  \notag \\ &\phantom{{}={}}\hspace{0.5cm}  -n_\beta a^\alpha) V^\mathbf{n}-n^\alpha
    (e^\delta_{d}\widetilde{E}_\beta^{d}-n^\delta n_\beta)\nabla_\delta V^\mathbf{n} \,.
 \end{align} 
From
 \begin{align}  
    \nabla_\beta G^\alpha&= e^\alpha_{c}\widetilde{E}_\gamma^{c} e^\delta_{d}\widetilde{E}_\beta^{d}\nabla_\delta\big(e^\gamma_{a}V^a\big)-e^\alpha_{c}\widetilde{E}_\gamma^{c} n^\delta n_\beta \nabla_\delta\big(e^\gamma_{a}V^a\big)  
     -n^\alpha n_\gamma e^\delta_{d}\widetilde{E}_\beta^{d}\nabla_\delta\big(e^\gamma_{a}V^a\big)
   \notag \\ &\phantom{{}={}}\hspace{0.5cm} +n^\alpha n_\gamma n^\delta n_\beta \nabla_\delta\big(e^\gamma_{a}V^a\big)-(e^\alpha_{a}K^a_{\ b}\widetilde{E}_\beta^{b}-n_\beta a^\alpha) V^\mathbf{n}-n^\alpha 
   e^\delta_{d}\widetilde{E}_\beta^{d} \nabla_\delta V^\mathbf{n}\notag \\ &\phantom{{}={}}\hspace{0.5cm} -n^\alpha n^\delta n_\beta \nabla_\delta V^\mathbf{n} \,, \notag 
 \end{align} 
integrating by parts the third and fourth terms, and recognizing geometrical quantities we arrive at
 \begin{align} \label{ecuacion_current}
 \nabla_\beta G^\alpha &= e^\alpha_{a}\widetilde{E}_\beta^{b}   \left(D_bV^a-K^a_{\ b}V^{\bf n}\right)
  -e^\alpha_{a}  n_\beta \left(   \widetilde{E}_\gamma^{a}n^\delta\nabla_\delta
   \big(e^\gamma_{c}V^c\big)-a^aV^{\bf n}\right)  \nonumber \\ &\phantom{{}={}}\hspace{0.5cm} -n^\alpha
  \widetilde{E}_\beta^{b}  \big(  D_bV^{\bf n}-K_{cb}V^c\big)+n^\alpha  
   n_\beta \big(  n^\delta\nabla_\delta V^\mathbf{n}-a_cV^c\big)    \,. 
\end{align}
We can go further by using the expression \eqref{liederivative} to write the projections
as
 \begin{align} \label{ecuacion_current}
 \nabla_\beta G^\alpha &= e^\alpha_{a}\widetilde{E}_\beta^{b}\big(D_b V^a-K^a_{\ b}V^{\bf n}\big) -e^\alpha_{a} 
  n_\beta \bigg( \frac{1}{N}\widetilde{E}_\gamma^{a}\mathcal{L}_m\big(e^\gamma_{c}V^c\big)+K^a_{\ c}V^c-a^aV^{\bf n}\bigg) 
    \nonumber \\ &\phantom{{}={}}\hspace{0.5cm} -n^\alpha
  \widetilde{E}_\beta^{b}  \big(D_bV^{\bf n}-K_{cb}V^c\big)+n^\alpha  
   n_\beta \bigg(\frac{1}{N}\mathcal{L}_m V^\mathbf{n}-a_cV^c\bigg)    \,. 
\end{align}
Taking this example as reference, we start the projections for the relevant terms in the like 
energy-momentum tensor~\eqref{Eq:equationst}.

Now, we start to decompose the background tensor 
 $t^{\mu\alpha\nu\beta}$
\begin{align}
  t^{\mu\alpha\nu\beta}&=e^\mu_{c}e^\alpha_{a}e^\nu_{d}e^\beta_{b}t^{cadb}-n^\mu e^\alpha_{a}
  e^\nu_{d}e^\beta_{b}t^{\mathbf{n}adb}-e^\mu_{c}n^\alpha e^\nu_{d}e^\beta_{b}t^{c\mathbf{n}db}
 -e^\mu_{c}e^\alpha_{a}n^\nu e^\beta_{b}t^{ca\mathbf{n}b}
   -e^\mu_{c}e^\alpha_{a}e^\nu_{d}n^\beta t^{cad\mathbf{n}} \notag  \\ &\phantom{{}={}}\hspace{0.5cm}+n^\mu e^\alpha_{a}n^\nu e^\beta_{b}t^{\mathbf{n}a\mathbf{n}b}
    +n^\mu e^\alpha_{a}e^\nu_{d}n^\beta t^{\mathbf{n}ad\mathbf{n}}+e^\mu_{c}n^\alpha n^\nu e^\beta_{b}t^{c\mathbf{n}\mathbf{n}b} 
    +e^\mu_{c}n^\alpha e^\nu_{d}n^\beta t^{c\mathbf{n}d\mathbf{n}} \,.
\end{align}
We apply the covariant derivative  
\begin{align}
  \nabla_\lambda t^{\mu\alpha\nu\beta}&=
  \nabla_\lambda\Big(e^\mu_{c}e^\alpha_{a}e^\nu_{d}e^\beta_{b}t^{cadb}
  -n^\mu e^\alpha_{a}e^\nu_{d}e^\beta_{b}t^{\mathbf{n}adb}-e^\mu_{c}n^\alpha e^\nu_{d}e^\beta_{b}t^{c\mathbf{n}db}  
   -e^\mu_{c}e^\alpha_{a}n^\nu e^\beta_{b}t^{ca\mathbf{n}b} \notag \\ &\phantom{{}={}}\hspace{0.5cm}
   -e^\mu_{c}e^\alpha_{a}e^\nu_{d}n^\beta t^{cad\mathbf{n}} +n^\mu e^\alpha_{a}n^\nu e^\beta_{b}t^{\mathbf{n}a\mathbf{n}b}
   +n^\mu e^\alpha_{a}e^\nu_{d}n^\beta t^{\mathbf{n}ad\mathbf{n}}+e^\mu_{c}n^\alpha n^\nu e^\beta_{b}t^{c\mathbf{n}\mathbf{n}b}
  \notag  \\ &\phantom{{}={}}\hspace{0.5cm}+e^\mu_{c}n^\alpha e^\nu_{d}n^\beta t^{c\mathbf{n}d\mathbf{n}}\Big) \,,
\end{align}
which acts over the $n^\mu$ four-vector as
\begin{align}
\nabla_\lambda t^{\mu\alpha\nu\beta} &=\nabla_\lambda\big(e^\mu_{c}e^\alpha_{a}e^\nu_{d}e^\beta_{b}t^{cadb}\big)-(\nabla_\lambda n^\mu) e^\alpha_{a}e^\nu_{d}e^\beta_{b}t^{\mathbf{n}adb}-n^\mu \nabla_\lambda \big(e^\alpha_{a}e^\nu_{d}e^\beta_{b}t^{\mathbf{n}adb}\big)
   \\ & -(\nabla_\lambda n^\alpha) e^\mu_{c} e^\nu_{d}e^\beta_{b}t^{c\mathbf{n}db}-n^\alpha \nabla_\lambda\big(e^\mu_{c} e^\nu_{d}e^\beta_{b}t^{c\mathbf{n}db}\big)-(\nabla_\lambda n^\nu)e^\mu_{c}e^\alpha_{a} e^\beta_{b}t^{ca\mathbf{n}b}\nonumber \\
&-n^\nu\nabla_\lambda\big(e^\mu_{c}e^\alpha_{a} e^\beta_{b}t^{ca\mathbf{n}b}\big)
   -(\nabla_\lambda n^\beta)e^\mu_{c}e^\alpha_{a}e^\nu_{d} t^{cad\mathbf{n}}-n^\beta \nabla_\lambda\big(e^\mu_{c}e^\alpha_{a}e^\nu_{d} t^{cad\mathbf{n}}\big) \notag \\
   & +(\nabla_\lambda n^\mu n^\nu+n^\mu \nabla_\lambda n^\nu) e^\alpha_{a} e^\beta_{b}t^{\mathbf{n}a\mathbf{n}b}+n^\mu n^\nu \nabla_\lambda\big(e^\alpha_{a} e^\beta_{b}t^{\mathbf{n}a\mathbf{n}b}\big) \notag \\
   &+(\nabla_\lambda n^\mu n^\beta+n^\mu \nabla_\lambda n^\beta) e^\alpha_{a}e^\nu_{d} t^{\mathbf{n}ad\mathbf{n}}+n^\mu n^\beta \nabla_\lambda\big(e^\alpha_{a}e^\nu_{d} t^{\mathbf{n}ad\mathbf{n}}\big) \notag \\
   &+(\nabla_\lambda n^\alpha n^\nu+n^\alpha \nabla_\lambda n^\nu) e^\mu_{c} e^\beta_{b}t^{c\mathbf{n}\mathbf{n}b}+n^\alpha n^\nu \nabla_\lambda\big(e^\mu_{c} e^\beta_{b}t^{c\mathbf{n}\mathbf{n}b}\big) \notag \\
   &+(\nabla_\lambda n^\alpha n^\beta+n^\alpha \nabla_\lambda n^\beta) e^\mu_{c} e^\nu_{d} t^{c\mathbf{n}d\mathbf{n}}+n^\alpha n^\beta \nabla_\lambda \big( e^\mu_{c} e^\nu_{d} t^{c\mathbf{n}d\mathbf{n}}\big) \notag\,.
\end{align}
Taking the contraction in the $\beta$ and $\lambda$ indices produces
\begin{align}
\nabla_\beta t^{\mu\alpha\nu\beta} &=\nabla_\beta\big(e^\mu_{c}e^\alpha_{a}e^\nu_{d}e^\beta_{b}t^{cadb}\big)-e^\mu_c  e^\alpha_{a}e^\nu_{d}K^c_{\ b}t^{\mathbf{n}adb}-n^\mu \nabla_\beta \big(e^\alpha_{a}e^\nu_{d}e^\beta_{b}t^{\mathbf{n}adb}\big) \notag \\ & -e^\alpha_a e^\mu_{c} e^\nu_{d}K^a_{\ b}t^{c\mathbf{n}db}-n^\alpha \nabla_\beta\big(e^\mu_{c} e^\nu_{d}e^\beta_{b}t^{c\mathbf{n}db}\big)-e^\nu_{d}e^\mu_{c}e^\alpha_{a} K^d_{\ b}t^{ca\mathbf{n}b}\nonumber \\
&-n^\nu\nabla_\beta\big(e^\mu_{c}e^\alpha_{a} e^\beta_{b}t^{ca\mathbf{n}b}\big)
   -e^\mu_{c}e^\alpha_{a}e^\nu_{d} K t^{cad\mathbf{n}}-n^\beta \nabla_\beta\big(e^\mu_{c}e^\alpha_{a}e^\nu_{d} t^{cad\mathbf{n}}\big) \notag \\
   & +(e^\mu_c K^c_{\ b} n^\nu+ n^\mu e^\nu_d K^d_{\ b}) e^\alpha_{a} t^{\mathbf{n}a\mathbf{n}b}+n^\mu n^\nu \nabla_\beta\big(e^\alpha_{a} e^\beta_{b}t^{\mathbf{n}a\mathbf{n}b}\big) \notag \\
   &+(e^\mu_c a^c +n^\mu K) e^\alpha_{a}e^\nu_{d} t^{\mathbf{n}ad\mathbf{n}}+n^\mu n^\beta \nabla_\beta\big(e^\alpha_{a}e^\nu_{d} t^{\mathbf{n}ad\mathbf{n}}\big) \notag \\
   &+(e^\alpha_a K^a_{\ b}n^\nu+n^\alpha e^\nu_d K^d_{\ b}) e^\mu_{c} t^{c\mathbf{n}\mathbf{n}b}+n^\alpha n^\nu \nabla_\beta\big(e^\mu_{c} e^\beta_{b}t^{c\mathbf{n}\mathbf{n}b}\big) \notag \\
   &+(e^\alpha_a a^a+n^\alpha K) e^\mu_{c} e^\nu_{d} t^{c\mathbf{n}d\mathbf{n}}+n^\alpha n^\beta \nabla_\beta \big( e^\mu_{c} e^\nu_{d} t^{c\mathbf{n}d\mathbf{n}}\big)\,.
\end{align}
Where we have used the expression for $a^\alpha$ and the extrinsic curvature $K_{ab}$. We continue by introducing Kronecker 
deltas in the terms with unprojected indices. Starting with the first term
\begin{align}
   \nabla_\beta\big(e^\mu_{c}e^\alpha_{a}e^\nu_{d}e^\beta_{b}t^{cadb}\big)&=  \delta^\mu_\rho \delta^\nu_\sigma \delta^\alpha_\lambda\delta^\beta_\kappa \nabla_\beta\big(e^\rho_{k}e^\lambda_{i}e^\sigma_{l}e^\kappa_{j}t^{kilj}\big) \,.
   \end{align}
Expanding the expression, using integration by parts  and identifying the recognizable quantities we obtain
\begin{align}
  \nabla_\beta\big(e^\mu_{c}e^\alpha_{a}e^\nu_{d}e^\beta_{b}t^{cadb}\big)  &=  e^\mu_a e^\nu_b 
  e^\alpha_c D_dt^{acbd}+n^\mu   e^\nu_b  e^\alpha_c   K_{kd}t^{kcbd} +e^\mu_a  n^\nu  e^\alpha_c    K_{ld}t^{acld}  \notag \\
   &\phantom{{}={}}\hspace{0.5cm}+e^\mu_a  e^\nu_b  n^\alpha       K_{id}t^{aibd}+e^\mu_a  e^\nu_b  e^\alpha_c  a_j t^{acbj}  \,,
\end{align}
Due to the skew symmetry of $t^{acbd}$ and the symmetry 
of $K_{ab}$ we can discard the third term. We obtain
\begin{align}
  \nabla_\beta\big(e^\mu_{c}e^\alpha_{a}e^\nu_{d}e^\beta_{b}t^{cadb}\big)  &=  e^\mu_a e^\nu_b
   e^\alpha_c \big(D_dt^{acbd}+ a_d t^{acbd} \big)+n^\mu   e^\nu_b  e^\alpha_c   \big(K_{ad}t^{acbd}  \big) \notag \\ 
   &\phantom{{}={}}\hspace{0.5cm} +e^\mu_a  e^\nu_b  n^\alpha       \big(K_{cd}t^{acbd}\big) \,,
\end{align}
Proceeding in the same way we obtain the following projections
\begin{align}
    \nabla_\beta \big(e^\alpha_{a}e^\nu_{d}e^\beta_{b}t^{\mathbf{n}adb}\big)&=  
    e^\nu_b e^\alpha_c\big(D_d t^{\mathbf{n} cbd} +a_d t^{\mathbf{n}cbd}\big)+e^\nu_b n^\alpha   \big(K_{cd}t^{\mathbf{n}cbd}\big)\,,
\\
  n^\beta \nabla_\beta\big(e^\mu_{c}e^\alpha_{a}e^\nu_{d} t^{cad\mathbf{n}}\big)&= e^\mu_a   e^\nu_b   e^\alpha_c \big( \widetilde{E}_\rho^a\widetilde{E}_\sigma^b\widetilde{E}_\lambda^c n^\beta \nabla_\beta\big(e^\rho_{i}e^\lambda_{k}e^\sigma_{j} t^{ikj\mathbf{n}}\big)\big) +n^\mu    e^\nu_b e^\alpha_c \big( a_a  t^{acb\mathbf{n}}\big)
  \notag \\
  &\phantom{{}={}}\hspace{0.5cm}+e^\mu_a  n^\nu e^\alpha_c   \big(  a_bt^{acb\mathbf{n}}\big)+e^\mu_a   e^\nu_b n^\alpha \big( a_c t^{acb\mathbf{n}}\big)\,,
\\
   \nabla_\beta\big(e^\alpha_{a} e^\beta_{b}t^{\mathbf{n}a\mathbf{n}b}\big)&=  e^\alpha_c \big(D_d t^{\mathbf{ n} c \bf n d}+a_dt^{\mathbf{n}c\mathbf{n}d}\big)+n^\alpha    \big(K_{kd}t^{\mathbf{n}k\mathbf{n}d}\big)\,,
\\
    n^\beta \nabla_\beta \big( e^\mu_{c} e^\nu_{d} t^{c\mathbf{n}d\mathbf{n}}\big)&=e^\mu_a   e^\nu_b \big(\widetilde{E}_\rho^a\widetilde{E}_\sigma^b  n^\beta \nabla_\beta \big( e^\rho_{i} e^\sigma_{j} t^{i\mathbf{n}j\mathbf{n}}\big)\big)+n^\mu    e^\nu_b \big( a_a   t^{a\mathbf{n}b\mathbf{n}}\big)  \notag \\
    &\phantom{{}={}}\hspace{0.5cm}+e^\mu_a   n^\nu  \big(  a_b   t^{a\mathbf{n}b\mathbf{n}} \big)\,.
\end{align}
Replacing the projections in the covariant divergence of the $t$ tensor we obtain a final expression for their projection
\begin{align}
\nabla_\beta t^{\mu\alpha\nu\beta}&=e^\mu_{\ a}e^\alpha_{\ c}e^\nu_{\ b}\big[\Theta^{acb}\big]-n^\mu e^\alpha_{\ c}e^\nu_{\ b}\big[\Sigma^{cb}\big]-e^\mu_{\ a}n^\alpha e^\nu_{\ b}\big[-\Sigma^{ab}\big]
    \notag \\ & -e^\mu_{\ a}e^\alpha_{\ c}n^\nu \big[\Phi^{ac}\big]+n^\mu e^\alpha_{\ c}n^\nu \big[\Lambda^c\big]+e^\mu_{\ a}n^\alpha n^\nu \big[-\Lambda^a\big] \,, 
\end{align}
where we defined the projections as
\begin{align}
    \Theta^{acb}&:= D_d t^{acbd}-K^a_{\ d}t^{\mathbf{n}cbd}-K^c_{\  d}t^{a\mathbf{n}bd} -K^b_{\ d}t^{ac\mathbf{n}d}-Kt^{acb\mathbf{n}} 
      \notag \\ &\phantom{{}={}}\hspace{0.5cm}  -\widetilde{E}_\rho^{a}\widetilde{E}_\lambda^{c}\widetilde{E}_\sigma^{b}n^\kappa \nabla_\kappa(e^\rho_{i}e^\lambda_{k}e^\sigma_{j} t^{ikj\mathbf{n}}) + a_dt^{acbd}+ a^at^{\mathbf{n}cb\mathbf{n}}+a^ct^{a\mathbf{n}b\mathbf{n}}   \,, 
\\
    \Sigma^{cb}&:=D_d t^{\mathbf{n}cbd}-K_{ad}t^{acbd}- K^b_{\ d} t^{\mathbf{n}c\mathbf{n}d}-  Kt^{\mathbf{n}cb\mathbf{n}}  - \widetilde{E}_\lambda^{c}\widetilde{E}_\sigma^{b}  n^\kappa \nabla_\kappa(e^\lambda_{k}e^\sigma_{j} t^{\mathbf{n}kj\mathbf{n}}) \notag \\ &\phantom{{}={}}\hspace{0.5cm}+    a_dt^{\mathbf{n}cbd}+ a_at^{acb\mathbf{n}} \,,
    \\
\Phi^{ac}&:= D_dt^{ac\mathbf{n}d} - K^a_{\ d}t^{\mathbf{n}c\mathbf{n}d}-K^c_{\ d}t^{a\mathbf{n}\mathbf{n}d}\,,
\\
\Lambda^c&:=D_dt^{\mathbf{n}c\mathbf{n}d}- K_{ad}t^{ac\mathbf{n}d} \,.
\end{align}

We continue by acting with an second covariant derivative
\begin{align}
    \nabla_\lambda\nabla_\beta t^{\mu\alpha\nu\beta}&=\nabla_\lambda\big(e^\mu_{a}
   e^\alpha_{c}e^\nu_{b}\Theta^{acb}\big)-(\nabla_\lambda n^\mu ) 
   e^\alpha_{c}e^\nu_{b}\Sigma^{cb}   -n^\mu \nabla_\lambda\big(e^\alpha_{c}e^\nu_{b}\Sigma^{cb}\big)
    \nonumber \\
    &+(\nabla_\lambda n^\alpha)e^\mu_{a} e^\nu_{b}\Sigma^{ab}+n^\alpha\nabla_\lambda\big(e^\mu_{a} e^\nu_{b}\Sigma^{ab}\big)-(\nabla_\lambda n^\nu)e^\mu_{\ a}e^\alpha_{c} \Phi^{ac} \notag \\ &-n^\nu\nabla_\lambda\big(e^\mu_{a}e^\alpha_{c} \Phi^{ac}\big)
  +\big(n^\mu \nabla_\lambda n^\nu+n^\nu\nabla_\lambda n^\mu \big) e^\alpha_{c} \Lambda^c
  +n^\mu n^\nu \nabla_\lambda\big(e^\alpha_{c} \Lambda^c\big)\nonumber \\ & -\big(n^\alpha \nabla_\lambda n^\nu+n^\nu\nabla_\lambda n^\alpha \big)e^\mu_{a} \Lambda^a -n^\alpha n^\nu\nabla_\lambda \big(e^\mu_{a} \Lambda^a\big) \,.
\end{align}
Contracting in the $\lambda$ and $\alpha$ indices we obtain
\begin{align}
    \nabla_\alpha \nabla_\beta t^{\mu\alpha\nu\beta}&=\nabla_\alpha\big(e^\mu_{a}e^\alpha_{c}e^\nu_{b}\Theta^{acb}\big)-n^\mu \nabla_\alpha\big(e^\alpha_{c}e^\nu_{b}\Sigma^{cb}\big)
    +n^\alpha\nabla_\alpha\big(e^\mu_{a} e^\nu_{b}\Sigma^{ab}\big)\notag \\ &-e^\mu_{a}e^\nu_{b}\big(K^a_{\ c}\Sigma^{cb}-K\Sigma^{ab}+K^b_{\ \ c} \Phi^{ac}+a^b \Lambda^a\big) -n^\nu\nabla_\alpha\big(e^\mu_{a}e^\alpha_{c} \Phi^{ac}\big)\notag \\ &+n^\mu e^\nu_{b}K^b_{\ c}\Lambda^c+e^\mu_{a} n^\nu \big(K^a_{\ c}  \Lambda^c -K\Lambda^a\big)+n^\mu n^\nu \nabla_\alpha\big(e^\alpha_{\ c} \Lambda^c\big)\notag \\ &-n^\nu n^\alpha \nabla_\alpha \big(e^\mu_{\ a} \Lambda^a\big)  \,.
\end{align}
We proceed as usual introducing Kronecker deltas in the unprojected terms, integrating by parts and recognizing 
 geometrical quantities. We list the projections obtained
\begin{align}
    \nabla_\alpha\big(e^\mu_{a}e^\alpha_{c}e^\nu_{b}\Theta^{acb}\big)&=(e^\mu_a \widetilde{E}_\rho^a -n^\mu n_\rho) (e^\nu_b\widetilde{E}_\sigma^b -n^\nu n_\sigma) (e^\alpha_c\widetilde{E}_\lambda^c-n^\alpha n_\lambda)\nabla_\alpha\big(e^\rho_{i}e^\lambda_{k}e^\sigma_{j}\Theta^{ikj}\big) \notag \\
    &=e^\mu_a e^\nu_b\big[D_c\Theta^{acb}+a_c\Theta^{acb}\big]+n^\mu   e^\nu_bK_{ac}\Theta^{acb}+e^\mu_a  n^\nu   K_{bc}\Theta^{acb}\,,
\end{align}
\begin{align}
    \nabla_\alpha\big(e^\alpha_c e^\nu_b\Sigma^{cb}\big)&= (e^\nu_b\widetilde{E}_\sigma^b -n^\nu n_\sigma) (e^\alpha_c\widetilde{E}_\lambda^c-n^\alpha n_\lambda)\nabla_\alpha\big(e^\lambda_k e^\sigma_j\Sigma^{kj}\big) \notag \\
    &= e^\nu_b \big[D_c\Sigma^{cb}+a_c\Sigma^{cb}\big]+n^\nu    K_{bc} \Sigma^{cb}\,,
\end{align}
\begin{align}
    n^\alpha\nabla_\alpha\big(e^\mu_a e^\nu_b\Sigma^{ab}\big)&=(e^\mu_a \widetilde{E}_\rho^a -n^\mu n_\rho) (e^\nu_b\widetilde{E}_\sigma^b -n^\nu n_\sigma)n^\alpha\nabla_\alpha\big(e^\rho_i e^\sigma_j\Sigma^{ij}\big) \notag \\
    &=e^\mu_a e^\nu_b\big[\widetilde{E}_\rho^a \widetilde{E}_\sigma^b n^\alpha\nabla_\alpha\big(e^\rho_i e^\sigma_j\Sigma^{ij}\big)\big] +n^\mu   e^\nu_b a_a\Sigma^{ab}+e^\mu_a  n^\nu  a_b\Sigma^{ab} \,,
\end{align}
\begin{align}
    \nabla_\alpha\big(e^\mu_a e^\alpha_c\Phi^{ac}\big)&= (e^\mu_a \widetilde{E}_\rho^a -n^\mu n_\rho)  (e^\alpha_c\widetilde{E}_\lambda^c-n^\alpha n_\lambda)\nabla_\alpha\big(e^\rho_i e^\lambda_k\Phi^{ik}\big) \notag \\
    &= e^\mu_a  \big[D_c\Phi^{ac}+a_c\Phi^{ac}\big]+n^\mu    K_{ac}\Phi^{ac}\,,
\end{align}
\begin{align}
    \nabla_\alpha\big(e^\alpha_c\Lambda^c\big)&=(e^\alpha_c\widetilde{E}_\lambda^c-n^\alpha n_\lambda)\nabla_\alpha\big(e^\lambda_k\Lambda^k\big)=D_c\Lambda^c+a_c\Lambda^c\,,
\end{align}
\begin{align}
    n^\alpha \nabla_\alpha\big(e^\mu_a\Lambda^a\big)&=(e^\mu_a \widetilde{E}_\rho^a -n^\mu n_\rho)  n^\alpha \nabla_\alpha\big(e^\rho_i\Lambda^i\big) 
    = e^\mu_a \big[\widetilde{E}_\rho^a n^\alpha \nabla_\alpha\big(e^\rho_i\Lambda^i\big)\big]+n^\mu  a_a\Lambda^a  \,.
\end{align}
Replacing the previous result in the expression for the second derivative we obtain
\begin{align}
    \nabla_\alpha \nabla_\beta t^{\mu\alpha\nu\beta}&=e^\mu_{a}e^\nu_{b}\big[D_c\Theta^{acb}+a_c\Theta^{acb}+\widetilde{E}_\rho^{a}\widetilde{E}_\sigma^{b} n^\alpha\nabla_\alpha\big(e^\rho_{c} e^\sigma_{d}\Sigma^{cd}\big)-K^a_{\ c}\Sigma^{cb}\notag \\
    &+K\Sigma^{ab}-K^b_{\ c} \Phi^{ac}-a^b \Lambda^a\big] -n^\mu e^\nu_{b}\big[D_c\Sigma^{cb}-  K_{ac}\Theta^{acb}-  K^b_{\ a}\Lambda^a\big]\nonumber \\
    &-e^\mu_{a}n^\nu \big[D_b\Phi^{ab}- K_{bc}\Theta^{acb}- a_b\Sigma^{ab} +a_b\Phi^{ab}-K^a_{\ b}  \Lambda^b +K\Lambda^a\notag \\
    &+\widetilde{E}_\rho^{a} n^\alpha \nabla_\alpha \big(e^\rho_{c} \Lambda^c\big)\big]+n^\mu n^\nu  \big[D_a \Lambda^a -    K_{ab}\Sigma^{ab} -  K_{ab} \Phi^{ab}\big] \,.
    \end{align}
Using the definitions of $\Theta^{acb},\Sigma^{cb},\Phi^{ac}$ and $\Lambda^c$ 
we obtain an explicit expression for the projection of the double covariant divergence of $t$
\begin{align}
    \nabla_\alpha\nabla_\beta t^{\mu\alpha\nu\beta}&=
    e^\mu_a e^\nu_b \mathcal{S}^{ab}-e^\mu_a 
    n^\nu \mathcal{S}^{a\mathbf{n}}-n^\mu e^\nu_b \mathcal{S}^{\mathbf{n}b}+n^\mu n^\nu \mathcal{S}^{\mathbf{n}\mathbf{n}}\,,
\end{align}
where
\begin{align}
    \mathcal{S}^{ab}&=D_c\Big(D_d t^{acbd}-K^a_{\ d}t^{\mathbf{n}cbd}-K^c_{\  d}t^{a\mathbf{n}bd}-K^b_{\ d}t^{ac\mathbf{n}d}-Kt^{acb\mathbf{n}}-\widetilde{E}_\mu^{a}\widetilde{E}_\alpha^{c}\widetilde{E}_\nu^{b}n^\beta \nabla_\beta(e^\mu_{i}e^\alpha_{k}e^\nu_{j} t^{ikj\mathbf{n}})\nonumber \\
 & + a_dt^{acbd}+ a^at^{\mathbf{n}cb\mathbf{n}}+a^ct^{a\mathbf{n}b\mathbf{n}} \Big) +a_c\Big(D_d t^{acbd}-K^a_{\ d}t^{\mathbf{n}cbd}-K^c_{\  d}t^{a\mathbf{n}bd}-K^b_{\ d}t^{ac\mathbf{n}d}-Kt^{acb\mathbf{n}}     \nonumber \\
 &   -\widetilde{E}_\mu^{a}\widetilde{E}_\alpha^{c}\widetilde{E}_\nu^{b}n^\beta \nabla_\beta(e^\mu_{i}e^\alpha_{k}e^\nu_{j} t^{ikj\mathbf{n}})
 + a_dt^{acbd}+ a^at^{\mathbf{n}cb\mathbf{n}}+a^ct^{a\mathbf{n}b\mathbf{n}} \Big)-\widetilde{E}_\mu^{a}\widetilde{E}_\nu^{b} n^\alpha\nabla_\alpha\Big(e^\mu_{i} e^\nu_{j}\big(D_l t^{i\mathbf{n}jl}     \nonumber \\
 &  -K_{kl}t^{ikjl}  - K^j_{\ l} t^{i\mathbf{n}\mathbf{n}l}-  Kt^{i\mathbf{n}j\mathbf{n}} - \widetilde{E}_\rho^{i}\widetilde{E}_\sigma^{j}  n^\beta \nabla_\beta(e^\rho_{c}e^\sigma_{d} t^{c\mathbf{n}d\mathbf{n}})+    a_lt^{i\mathbf{n}jl}+ a_kt^{ikj\mathbf{n}}\big)\Big)\nonumber \\
 &-K^a_{\ c}\Big(D_d t^{\mathbf{n}cbd}-K_{ed}t^{ecbd}- K^b_{\ d} t^{\mathbf{n}c\mathbf{n}d}-  Kt^{\mathbf{n}cb\mathbf{n}} - \widetilde{E}_\alpha^{c}\widetilde{E}_\nu^{b}  n^\beta \nabla_\beta(e^\alpha_{k}e^\nu_{j} t^{\mathbf{n}kj\mathbf{n}})+    a_dt^{\mathbf{n}cbd}+ a_at^{acb\mathbf{n}}\Big)\nonumber \\
 & -K\Big(D_d t^{a\mathbf{n}bd}-K_{cd}t^{acbd}- K^b_{\ d} t^{a\mathbf{n}\mathbf{n}d}-  Kt^{a\mathbf{n}b\mathbf{n}} - \widetilde{E}_\alpha^{a}\widetilde{E}_\nu^{b}  n^\beta \nabla_\beta(e^\alpha_{\ i}e^\nu_{\ j} t^{i\mathbf{n}j\mathbf{n}})+    a_dt^{a\mathbf{n}bd}+ a_ct^{acb\mathbf{n}}\Big)\nonumber \\
 & -K^b_{\ c} \Big(D_dt^{ac\mathbf{n}d} - K^a_{\ d}t^{\mathbf{n}c\mathbf{n}d}-K^c_{\ d}t^{a\mathbf{n}\mathbf{n}d}\Big)+a^b \Big(D_dt^{\mathbf{n}ad\mathbf{n}}- K_{cd}t^{cad\mathbf{n}} \Big)\,,
\end{align}
\begin{align}
   \mathcal{S}^{a\mathbf{n}}&=D_c\Big(D_dt^{ac\mathbf{n}d} - K^a_{\ d}t^{\mathbf{n}c\mathbf{n}d}-K^c_{\ d}t^{a\mathbf{n}\mathbf{n}d}\Big)- K_{bc}\Big(D_d t^{acbd}-K^a_{\ d}t^{\mathbf{n}cbj}-K^c_{\  d}t^{a\mathbf{n}bd}-K^b_{\ d}t^{ac\mathbf{n}d} \nonumber \\
    &  -Kt^{acb\mathbf{n}}-\widetilde{E}_\mu^{a}\widetilde{E}_\alpha^{c}\widetilde{E}_\nu^{b}n^\beta \nabla_\beta(e^\mu_{i}e^\alpha_{k}e^\nu_{j} t^{ikj\mathbf{n}})  + a_jt^{ailj}+ a^at^{\mathbf{n}il\mathbf{n}}+a^it^{a\mathbf{n}l\mathbf{n}} \Big)\nonumber \\
 & -K^a_{\ c}  \Big(D_dt^{\mathbf{n}c\mathbf{n}d}- K_{ed}t^{ec\mathbf{n}d} \Big) -K\Big(D_dt^{a\mathbf{n}\mathbf{n}d}- K_{cd}t^{ac\mathbf{n}d} \Big)\nonumber \\
 & + a_b\Big(D_d t^{a\mathbf{n}bd}-K_{cd}t^{acbd}- K^b_{\ \ d} t^{a\mathbf{n}\mathbf{n}d}-  Kt^{a\mathbf{n}b\mathbf{n}} - \widetilde{E}_\alpha^{a}\widetilde{E}_\nu^{b}  n^\beta \nabla_\beta(e^\alpha_{c}e^\nu_{d} t^{c\mathbf{n}d\mathbf{n}})+    a_dt^{a\mathbf{n}bd}+ a_ct^{acb\mathbf{n}}\Big) \nonumber \\
 & +a_c\Big(D_dt^{ac\mathbf{n}d} - K^a_{\ d}t^{\mathbf{n}c\mathbf{n}d}-K^c_{\ d}t^{a\mathbf{n}\mathbf{n}d}\Big)-\widetilde{E}_\mu^{a} n^\alpha \nabla_\alpha \Big(e^\mu_{i} \big(D_dt^{i\mathbf{n}\mathbf{n}d}- K_{cd}t^{ic\mathbf{n}d}\big)\Big)\,,
\end{align}
\begin{align}
   \mathcal{S}^{\mathbf{n}b}&=D_c\Big(D_d t^{\mathbf{n}cbd}-K_{ad}t^{acbd}- K^b_{\ d} t^{\mathbf{n}c\mathbf{n}d}-  Kt^{\mathbf{n}cb\mathbf{n}} - \widetilde{E}_\alpha^{c}\widetilde{E}_\nu^{b}  n^\beta \nabla_\beta(e^\alpha_{k}e^\nu_{j} t^{\mathbf{n}kj\mathbf{n}})\nonumber \\
 &+    a_dt^{\mathbf{n}cbd}+ a_at^{acb\mathbf{n}}\Big) -  K^b_{\ c}\Big(D_dt^{\mathbf{n}c\mathbf{n}d}- K_{aj}t^{ac\mathbf{n}d} \Big)\,,
\end{align}
\begin{align}
    \mathcal{S}^{\mathbf{n}\mathbf{n}}&=D_c\Big(D_dt^{\mathbf{n}c\mathbf{n}d}- K_{ad}t^{ac\mathbf{n}d} \Big)  -    K_{cb}\Big(D_d t^{\mathbf{n}cbd}-K_{ad}t^{acbd}- K^b_{\ d} t^{\mathbf{n}c\mathbf{n}d}
    -  Kt^{\mathbf{n}cb\mathbf{n}}\nonumber \\
    & - \widetilde{E}_\alpha^{c}\widetilde{E}_\nu^{\ b}  n^\beta \nabla_\beta(e^\alpha_{k}e^\nu_{j} t^{\mathbf{n}kj\mathbf{n}})
    +    a_dt^{\mathbf{n}cbd}+ a_at^{acb\mathbf{n}}\Big)\,.
\end{align}
\section{Purely spatial $t$ and late acceleration: Computational details}\label{AppendixC}
With the aim to satisfy isotropy and homogeneity, we used in section \ref{SectionV} the ansatz \eqref{rel1}, which we recall here
\begin{eqnarray}\label{rel1b}
	q_{cd}t^{acbd}-\frac{1}{3}q^{ab}q_{ef}q_{cd}t^{ecfd}&=&0 \,.
\end{eqnarray} 
This means that we can make the first trace of $t^{acbd}$ (i.e. $q_{cd}t^{acbd}$) 
proportional to  its second traces times a metric $q^{ab}$.
Taking the derivative of the above equation and considering that $q_{ab}$ is a Friedmann metric we get that
\begin{eqnarray} 
	\label{rel1c}
	q_{cd}\dot{t}^{acbd}-\frac{1}{3}q^{ab}q_{ef}q_{cd}\dot{t}^{ecfd}&=&0\,.
\end{eqnarray}

Also, we consider~\eqref{mom_constraint} which setting it to zero corresponds to the momentum 
constraint
\begin{eqnarray}
	- 2 H q_{cd} D_b t^{a c d b}=0\,.
\end{eqnarray}
Using metric compatibility and \eqref{rel1b} this relation can be recast in
\begin{eqnarray}
	\frac{2}{3} H \partial^a (q_{b f} q_{c d} t^{b c f d})=0\,,
\end{eqnarray}
which defines that the double trace of $t$ is only a function of time
\begin{eqnarray}
	\label{doubletraceft}
	q_{ab}q_{cd}t^{acbd}=f(t)\,.
\end{eqnarray}
Considering the result \eqref{rel1c}, it is possible to show that
\begin{eqnarray}
	\label{rel1d}
	-\frac{1}{3}q^{ab}q_{ef}\Big( D_c D_d t^{ecfd}+ D_c D_d t^{fced}\Big)=0\,.
\end{eqnarray}
Thus, using \eqref{rel1b}, \eqref{rel1c} and \eqref{rel1d} in the tensor $ T_{\textrm{t-spacelike}}^{\langle ab \rangle}$ in \eqref{tperp0} we arrive at
\begin{eqnarray}
	 T_{\textrm{t-spacelike}}^{\langle ab \rangle}=(-1)  ( D_c D_d t^{acbd}+ D_c D_d t^{bcad})\,.
\end{eqnarray}
With the previous considerations, the tensor \eqref{TRtspatial} now is written as 
\begin{eqnarray}
	&&  (T_{\textrm{t-spacelike}}^{Rt})^{\mu\nu}= e^\mu_{ a}e^\nu_{ b}\bigg[ \bigg(\frac{5k}{a(t)^2}-2\dot{H}+H^2\bigg)f(t)-2H\dot{f}(t)\bigg]\frac{1}{3}q^{ab}
	\nonumber \\
	& & +n^\mu n^\nu\bigg[\bigg(\frac{k}{a(t)^2}+3H^2\bigg)f(t)\bigg]\,.
\end{eqnarray}

Note that the form of the tensor agrees well with the expected result based on 
maximally symmetric tensors and invariant tensor, custodial symmetry we found in the sections before section \ref{SectionV}.

Now, to obtain the explicit form of the components of $t$, we will solve using spherical coordinates $(r,\theta,\phi)$. To solve explicitly \eqref{rel1b}
we note that
\begin{eqnarray}
	q_{cd}t^{acbd}&=&   q_{rr}t^{arbr}+q_{\theta\theta}t^{a\theta b \theta}+q_{\phi\phi}t^{a\phi b\phi}  \,, \\
	q_{ab}q_{cd}t^{acbd}&=&  2\big(q_{rr} q_{\theta\theta}t^{r\theta r \theta}+q_{rr}q_{\phi\phi}t^{r\phi  r\phi}
	+q_{\theta\theta}q_{\phi\phi}t^{\theta \phi \theta \phi}\big)\,. \label{explicitdoubtr}
\end{eqnarray}
Thus, using these explicit forms in \eqref{rel1b} and after some arrangements, the six independent equations that arise give
\begin{eqnarray}
	\label{eqfort0}   t^{r\phi\theta\phi}=t^{r\theta\phi\theta}=t^{\theta r\phi r}=0\,,
\end{eqnarray}
and
\begin{eqnarray}
	q_{\theta\theta}t^{r\theta r\theta }+q_{\phi\phi}t^{r\phi r\phi}=2q^{rr}q_{\theta\theta}q_{\phi\phi}t^{\theta \phi \theta \phi}  \label{eqfort1} \,, \\
	q_{rr}t^{\theta r\theta r}+ q_{\phi\phi}t^{\theta \phi\theta \phi}=2q^{\theta\theta}q_{rr}q_{\phi\phi}t^{r\phi  r\phi}  \label{eqfort2}   \,, \\
	q_{rr}t^{\phi r\phi r}+q_{\theta\theta }t^{\phi \theta\phi \theta}=2q^{\phi\phi}q_{rr} q_{\theta\theta}t^{r\theta r \theta}\,. \label{eqfort3}
\end{eqnarray}
 
Arranging \eqref{eqfort3} in terms of $t^{r \theta r \theta} we have$
\begin{eqnarray}
	\frac{1}{2}q_{\phi\phi}\big(q^{\theta\theta} t^{\phi r\phi r}+q^{rr} t^{\phi \theta\phi \theta}\big)=t^{r\theta r \theta} \,.
\end{eqnarray}

And replacing this equation in the other two, we obtain the identical equations
\begin{eqnarray}
	t^{r\phi r\phi}=q^{rr}q_{\theta\theta}t^{\theta \phi \theta \phi} \,, \label{trfrrel}\\
	t^{\phi \theta\phi \theta}=q^{\theta\theta}q_{rr}t^{r\phi  r\phi} \,.
\end{eqnarray}
Therefore, writing in terms of $t^{\theta \phi \theta \phi}$ and the metric explicitly, the components of $t$ that we found are
\begin{eqnarray}\label{explicit_sol}
	t^{r\phi\theta\phi}&=&t^{r\theta\phi\theta}=t^{\theta r\phi r}=0 \,, \\
	t^{r\theta r \theta} &=& r^2 (1-kr^2)\sin^2\theta t^{\theta\phi  \theta\phi}\,, \\
	t^{r\phi r\phi}&=&r^2(1-kr^2) t^{\theta \phi \theta \phi} \,.
\end{eqnarray}

In order to get $t^{\theta \phi \theta \phi}$, and in consequence the explicit form for the rest, we use \eqref{doubletraceft} and \eqref{trfrrel} in \eqref{explicitdoubtr} and we have 
\begin{eqnarray}  \label{exp0}
	6a(t)^4 r^4\sin^2\theta t^{\theta\phi\theta\phi} =f(t)\,.
\end{eqnarray}

To gather all the time dependency in one function, we redefine $\eta=\frac{f}{6a^4}$ and we have
\begin{eqnarray}   t^{\theta\phi\theta\phi}&=&\frac{ \eta(t)}{r^4 \sin^2\theta }\,.
\end{eqnarray}

With this expression we got all the components of $t$ and they satisfy that
\begin{eqnarray}
	D_c D_d t^{acbd}+ D_c D_d t^{bcad}&=&0\,,
\end{eqnarray}
making $T_{\textrm{t-spacelike}}^{\langle ab \rangle}$.
\section{Projections in the bumblebee model   }\label{AppendixD}
Recall the energy-momentum tensor for the bumblebee field
\begin{eqnarray}
  (T_B)^{\mu\nu}&=& \kappa\bigg[2V' B^\mu B^\nu+B^\mu_{\ \kappa}B^{\nu\kappa} -\bigg( V+\frac{1}{4}B^{\lambda\kappa}B_{\lambda\kappa}  \bigg)g^{\mu\nu}\bigg]\nonumber  \\
  &\ &+  \frac{\xi}{2} \bigg[B^\lambda B^\kappa R_{\lambda\kappa}g^{\mu\nu}-2\big(g^{\mu\rho}B^\nu +g^{\nu\rho}B^\mu\big) B^\sigma R_{\rho\sigma} \nonumber \\
  &\ &+ \nabla_\lambda\nabla^\mu \big(B^\lambda B^\nu\big)+\nabla_\lambda\nabla^\nu \big(B^\lambda B^\mu\big) \nonumber \\
  &\ &-\nabla_\kappa\nabla_\lambda \big(B^\lambda B^\kappa \big)g^{\mu\nu} -\nabla_\lambda\nabla^\lambda\big( B^\mu B^\nu\big)   \bigg] \,.
\end{eqnarray}
We notice that the main projection to compute corresponds to the second covariant derivative for a product of two bumblebee fields with all free indexes. All the terms that involves second covariant derivatives can be obtained from this result.

We start decomposing the bumblebee field in their normal and tangential projections
\begin{equation}
    B^\alpha=e^\alpha_{a}B^a-n^\alpha B^\mathbf{n} \,.
\end{equation}
With this result the product of $B^\alpha$ fields have the following projections
\begin{eqnarray}
   B^\alpha B^\beta &=&e^\alpha_{a}e^\beta_{b}B^a B^b-n^\alpha e^\beta_{b} B^\mathbf{n} B^b -e^\alpha_{a}n^\beta B^a B^\mathbf{n}+n^\alpha n^\beta \big(B^\mathbf{n}\big)^2  \,.
\end{eqnarray}
Taking the first covariant derivative of the product
\begin{eqnarray}
  \nabla_\nu\big( B^\alpha B^\beta\big) &=&\nabla_\nu\big(e^\alpha_{a}e^\beta_{b}B^a B^b\big)-e^\alpha_{a} e^\beta_{b}\big(\widetilde{E}_\nu^{d}K^a_{\ d}- n_\nu a^a\big) B^\mathbf{n} B^b\nonumber \\
  &&-n^\alpha \nabla_\nu\big(e^\beta_{b} B^\mathbf{n} B^b\big) -e^\alpha_{a}e^\beta_{b}\big(\widetilde{E}_\nu^{d}K^b_{\ d}-n_\nu a^b \big)  B^a B^\mathbf{n}\nonumber \\
  &\ &-n^\beta \nabla_\nu\big(e^\alpha_{a} B^a B^\mathbf{n}\big)+ e^\alpha_{a}n^\beta\big(\widetilde{E}_\nu^{d}K^a_{\ d}- n_\nu a^a\big)\big(B^\mathbf{n}\big)^2 \nonumber \\
  &&+ n^\alpha e^\beta_{b}\big(\widetilde{E}_\nu^{d}K^b_{\ d}-n_\nu a^b \big)\big(B^\mathbf{n}\big)^2  +n^\alpha n^\beta \nabla_\nu\Big(\big(B^\mathbf{n}\big)^2 \Big) \,.
\end{eqnarray}
Introducing Kronecker deltas and identifying the geometric quantities we obtain the following projections:
\begin{eqnarray}
 \nabla_\nu\big(e^\alpha_{a}e^\beta_{b}B^a B^b\big)&=&e^\alpha_{\ a} e^\beta_{\ b} \widetilde{E}_\nu^{d}D_d\big(B^a B^b\big)+e^\alpha_{a} n^\beta \widetilde{E}_\nu^{d} K_{db}B^a B^b \nonumber \\
 &&+n^\alpha e^\beta_{b}\widetilde{E}_\nu^{d}K_{da}B^a B^b\nonumber \\
 &\ &-e^\alpha_{a} e^\beta_{b}n_\nu  \Big(\widetilde{E}_\gamma^{a} \widetilde{E}_\delta^{b}n^\lambda \nabla_\lambda\big(e^\gamma_{c}e^\delta_{d}B^c B^d\big)\Big)\nonumber \\
 &&-e^\alpha_{a}n^\beta  n_\nu a_b B^a B^b-n^\alpha e^\beta_{b}n_\nu a_aB^a B^b \,,
\end{eqnarray}
\begin{eqnarray}
\nabla_\nu\big(e^\beta_{\ b} B^\mathbf{n} B^b\big)&=&e^\beta_{b}\widetilde{E}_\nu^{d} D_d \big( B^\mathbf{n} B^b\big) +n^\beta \widetilde{E}_\nu^{d} K_{db}\big( B^\mathbf{n} B^b\big)\nonumber \\
&&-e^\beta_{b} n_\nu\Big(\widetilde{E}_\delta^{b}n^\lambda\nabla_\lambda\big(e^\delta_{d} B^\mathbf{n} B^d\big)\Big)-n^\beta n_\nu a_b B^\mathbf{n} B^b \,,
\end{eqnarray}
\begin{eqnarray}
\nabla_\nu\Big(\big(B^\mathbf{n}\big)^2 \Big)&=&e_\nu^{\ d}D_d \Big(\big(B^\mathbf{n}\big)^2 \Big)-n_\nu n^\lambda \nabla_\lambda\Big(\big(B^\mathbf{n}\big)^2 \Big) \,.
\end{eqnarray}
With this results we obtain the projection for the first covariant derivative
\begin{eqnarray}
    \nabla_\nu\big(B^\alpha B^\beta\big)&=&e^\alpha_{a} e^\beta_{b} 
    \widetilde{E}_\nu^{d}\Big[D_d\big(B^a B^b\big)-K^a_{\ d}B^\mathbf{n}B^b-K^b_{\ d}B^a B^\mathbf{n}\Big]\nonumber \\
    &&-e^\alpha_{a} n^\beta \widetilde{E}_\nu^{d} \Big[D_d \big(  B^a B^\mathbf{n}\big)-K_{db}B^a B^b-K^a_{\ d}\big(B^\mathbf{n}\big)^2\Big] \nonumber \\
    &&-n^\alpha e^\beta_{b}\widetilde{E}_\nu^{d}\Big[ D_d \big( B^\mathbf{n} B^b\big)-K_{da}B^a B^b-K^b_{\ d}\big(B^\mathbf{n}\big)^2\Big]\nonumber \\
 & &-e^\alpha_{a} e^\beta_{b}n_\nu  \Big[\widetilde{E}_\gamma^{a} \widetilde{E}_\delta^{b}n^\lambda \nabla_\lambda\big(e^\gamma_{c}e^\delta_{d}B^c B^d\big)
 -  a^a B^\mathbf{n} B^b- a^b   B^a B^\mathbf{n}\Big]\nonumber \\
 & &+e^\alpha_{a}n^\beta  n_\nu\Big[\widetilde{E}_\gamma^{a}n^\lambda\nabla_\lambda\big(e^\gamma_{c}  B^c B^\mathbf{n}\big)- 
 a_b B^a B^b-a^a \big(B^\mathbf{n}\big)^2\Big]   \notag \\ &&+n^\alpha e^\beta_{b}n_\nu \Big[\widetilde{E}_\delta^{b}n^\lambda\nabla_\lambda\big(e^\delta_{d} B^\mathbf{n} B^d\big)-a_aB^a B^b-a^b \big(B^\mathbf{n}\big)^2\Big]\nonumber \\
 & & +n^\alpha n^\beta \widetilde{E}_\nu^{d}\Big[D_d \Big(\big(B^\mathbf{n}\big)^2\Big)-K_{db}B^\mathbf{n}B^b-K_{da}B^a B^\mathbf{n} \Big]\nonumber \\
 & &-n^\alpha n^\beta n_\nu \Big[n^\lambda \nabla_\lambda\Big(\big(B^\mathbf{n}\big)^2 \Big)- a_b B^\mathbf{n} B^b- a_a  B^a B^\mathbf{n}\Big] 
\end{eqnarray}
We define the following quantities in order to simplify the second derivative expression
\begin{eqnarray}
    \Phi^{ab}_{\ \ d}&:=&D_d\big(B^a B^b\big)-K^a_{\ d}B^\mathbf{n}B^b-K^b_{\ d}B^a B^\mathbf{n} \,,\\
    \Theta^a_{\ d}&:=& D_d \big(  B^a B^\mathbf{n}\big)-K_{db}B^a B^b-K^a_{\ d}\big(B^\mathbf{n}\big)^2  \,, \\
    \Psi^{ab}&:=&\widetilde{E}_\gamma^{a} \widetilde{E}_\delta^{b}n^\lambda \nabla_\lambda\big(e^\gamma_{c}e^\delta_{d}B^c B^d\big)-  a^a B^\mathbf{n} B^b- a^b   B^a B^\mathbf{n} \,,\\
    \Pi^a &:=& \widetilde{E}_\gamma^{a}n^\lambda\nabla_\lambda\big(e^\gamma_{c}  B^c B^\mathbf{n}\big)- a_b B^a B^b-a^a \big(B^\mathbf{n}\big)^2 \,,\\
    \Sigma_d &:=&D_d \Big(\big(B^\mathbf{n}\big)^2\Big)-K_{db}B^\mathbf{n}B^b-K_{da}B^a B^\mathbf{n} \,,\\
    \Omega &:=& n^\lambda \nabla_\lambda\Big(\big(B^\mathbf{n}\big)^2 \Big)- a_b B^\mathbf{n} B^b- a_a  B^a B^\mathbf{n} \,,
\end{eqnarray}
obtaining
\begin{align}
      \nabla_\nu\big(B^\alpha B^\beta\big)&=e^\alpha_{a} e^\beta_{b} \widetilde{E}_\nu^{n}\Phi^{ab}_{\ \ d}
      -e^\alpha_{a} n^\beta \widetilde{E}_\nu^{n} \Theta^a_{\ d} -n^\alpha e^\beta_{b}\widetilde{E}_\nu^{d}\Theta^b_{\ d}\nonumber\\
      &-e^\alpha_{a} e^\beta_{b}n_\nu  \Psi^{ab}+e^\alpha_{a}n^\beta  n_\nu\Pi^a+n^\alpha e^\beta_{b}n_\nu \Pi^b\nonumber \\
 & +n^\alpha n^\beta \widetilde{E}_\nu^{d}\Sigma_d-n^\alpha n^\beta n_\nu \Omega 
\end{align}
We continue taking the second covariant derivative
\begin{eqnarray}
    \nabla_\mu  \nabla_\nu\big(B^\alpha B^\beta\big) &=&\nabla_\mu \big(e^\alpha_{a} e^\beta_{b} \widetilde{E}_\nu^{d}\Phi^{ab}_{\ \ d}\big)\nonumber \\
 & &-\big(\widetilde{E}_\mu^{c}K^b_{\ c}-a^b n_\mu\big) e^\alpha_{a}e^\beta_{b} \widetilde{E}_\nu^{\ d} \Theta^a_{\ d}-n^\beta\nabla_\mu\big(e^\alpha_{a}  \widetilde{E}_\nu^{d} \Theta^a_{\ d}\big)\nonumber \\
 & &-\big(\widetilde{E}_\mu^{c}K^a_{\ c}-a^a n_\mu\big)e^\alpha_{a} e^\beta_{b}\widetilde{E}_\nu^{d}\Theta^b_{\ d}-n^\alpha \nabla_\mu\big(e^\beta_{b}\widetilde{E}_\nu^{d}\Theta^b_{\ d}\big)\nonumber \\
    & &-\big(\widetilde{E}_\mu^{c}K_{cd}-a_d n_\mu\big) e^\alpha_{a} e^\beta_{b} \widetilde{E}_\nu^{d}  \Psi^{ab}-n_\nu\nabla_\mu\big(e^\alpha_{a} e^\beta_{b}  \Psi^{ab}\big)\nonumber \\
    & &+\Big[e^\beta_{b}\big(\widetilde{E}_\mu^{c}K^b_{\ c}-a^b n_\mu\big)  n_\nu+n^\beta \widetilde{E}_\nu^{d}\big(\widetilde{E}_\mu^{c}K_{cd}-a_d n_\mu\big)\Big]e^\alpha_{a}\Pi^a\nonumber \\
    &&+n^\beta  n_\nu\nabla_\mu\big(e^\alpha_{a}\Pi^a\big)\nonumber \\
 & &+\Big[e^\alpha_{a}\big(\widetilde{E}_\mu^{c}K^a_{\ c}-a^a n_\mu\big)  n_\nu+n^\alpha \widetilde{E}_\nu^{d}\big(\widetilde{E}_\mu^cK_{cd}-a_d n_\mu\big)\Big] e^\beta_{b}\Pi^b\nonumber \\
 &&+n^\alpha n_\nu \nabla_\mu\big(e^\beta_{b}\Pi^b \big)\nonumber \\
 & &+\Big[\big(\widetilde{E}_\mu^{c}K^a_{\ c}-a^a n_\mu\big)e^\alpha_{a} n^\beta+n^\alpha e^\beta_{b}\big(\widetilde{E}_\mu^{c}K^b_{\ c}-a^b n_\mu\big)\Big] \widetilde{E}_\nu^{d}\Sigma_d\nonumber \\
 &&+n^\alpha n^\beta \nabla_\mu\big( \widetilde{E}_\nu^{d}\Sigma_d\big)\nonumber \\
 & &-\Big[e^\alpha_{a}\big(\widetilde{E}_\mu^{c}K^a_{\ c}-a^a n_\mu\big) n^\beta n_\nu+n^\alpha e^\beta_{b}\big(\widetilde{E}_\mu^{c}K^b_{\ c}-a^b n_\mu\big) n_\nu\Big] \Omega\nonumber \\
 &&-n^\alpha n^\beta \widetilde{E}_\nu^{d}\big(\widetilde{E}_\mu^{c}K_{cd}-a_d n_\mu\big) \Omega -n^\alpha n^\beta n_\nu \nabla_\mu\Omega  \,.
\end{eqnarray}
As usual we introduce the deltas in the non-explicit terms and we recognize the geometrical quantities obtaining 
\begin{align}
 \nabla_\mu \big(e^\alpha_{a} e^\beta_{b} \widetilde{E}_\nu^{d}\Phi^{ab}_{\ \ d}\big)&=  \widetilde{E}_\mu^{c} e^\alpha_{a}e^\beta_{b}\widetilde{E}_\nu^{d}  D_c\Phi^{ab}_{\ \ d}- n_\mu  e^\alpha_{a}e^\beta_{b}\widetilde{E}_\nu^{d}\Big( \widetilde{E}_\gamma^{a}\widetilde{E}_\delta^{b} e^\lambda_{d} n^\kappa\nabla_\kappa \big(e^\gamma_{c} e^\delta_{e} \widetilde{E}_\lambda^{l}\Phi^{ce}_{\ \ l}\big)\Big)\nonumber \\
&+\widetilde{E}_\mu^{c} n^\alpha  e^\beta_{b}\widetilde{E}_\nu^{d}K_{ca}\Phi^{ab}_{\ \ d}+\widetilde{E}_\mu^{c} e^\alpha_{a}n^\beta \widetilde{E}_\nu^{d} K_{cb}\Phi^{ab}_{\ \ d}+\widetilde{E}_\mu^{c} e^\alpha_{a}e^\beta_{b} n_\nu K^d_{\ c}\Phi^{ab}_{\ \ d}\nonumber \\
&- n_\mu n^\alpha  e^\beta_{b} \widetilde{E}_\nu^{d}  a_a\Phi^{ab}_{\ \ d}- n_\mu  e^\alpha_{a}n^\beta \widetilde{E}_\nu^{d} a_b  \Phi^{ab}_{\ \ d}-n_\mu  e^\alpha_{a}e^\beta_{b} n_\nu   a^d  \Phi^{ab}_{\ \ d}  \,,
\end{align}

\begin{align}
 \nabla_\mu\big(e^\alpha_{a}\widetilde{E}_\nu^{d}\Theta^a_{\ d}\big)&= \widetilde{E}_\mu^{c} \widetilde{E}_\nu^{d} e^\alpha_{a}D_c\Theta^a_{\ d}- n_\mu  \widetilde{E}_\nu^{d} e^\alpha_{a}\Big(  e^\lambda_{d}\widetilde{E}_\gamma^{a}n^\kappa\nabla_\kappa\big(e^\gamma_{c}\widetilde{E}_\lambda^{l}\Theta^c_{\ l}\big)\Big)\nonumber \\
 &+\widetilde{E}_\mu^{c}  n_\nu  e^\alpha_{a} K^d_{\ c}\Theta^a_{\ d}+\widetilde{E}_\mu^{c} \widetilde{E}_\nu^{d} n^\alpha   K_{ca}\Theta^a_{\ d}- n_\mu  n_\nu  e^\alpha_{a}a^d\Theta^a_{\ d}- n_\mu \widetilde{E}_\nu^{d} n^\alpha a_a \Theta^a_{\ d}  \,,
 \end{align}
 \begin{align}
 \nabla_\mu\big(e^\alpha_{a}e^\beta_{b}\Psi^{ab}\big)&=\widetilde{E}_\mu^{c} e^\alpha_{a}e^\beta_{b}D_c\Psi^{ab}- n_\mu  e^\alpha_{a}  e^\beta_{b}\Big(  \widetilde{E}_\gamma^{a}\widetilde{E}_\delta^{b}n^\kappa \nabla_\kappa\big(e^\gamma_{c}e^\delta_{d}\Psi^{cd}\big)\Big)\nonumber \\
 &+\widetilde{E}_\mu^{c} n^\alpha  e^\beta_{b} K_{ca}\Psi^{ab}+\widetilde{E}_\mu^{c} e^\alpha_{a}  n^\beta K_{cb}\Psi^{ab}\nonumber \\
 &- n_\mu  e^\alpha_{a}  n^\beta  a_b\Psi^{ab} - n_\mu n^\alpha   e^\beta_{b} a_a\Psi^{ab}  \,,
\end{align}
\begin{eqnarray}
    \nabla_\mu\big(e^\alpha_{a}\Pi^a\big)&=&   \widetilde{E}_\mu^{c} e^\alpha_{a} D_c\Pi^a- n_\mu  e^\alpha_{a}\Big(\widetilde{E}_\gamma^{a}n^\kappa \nabla_\kappa\big(e^\gamma_{c}\Pi^c\big)\Big)\nonumber \\
     &&+\widetilde{E}_\mu^{c} n^\alpha  K_{ca}\Pi^a- n_\mu n^\alpha a_a\Pi^a \,,
     \end{eqnarray}
     \begin{eqnarray}
 \nabla_\mu\big(\widetilde{E}_\nu^{d}\Sigma_d\big)&=&\widetilde{E}_\mu^{c} \widetilde{E}_\nu^{d}D_c\Sigma_d- n_\mu  \widetilde{E}_\nu^{d}\Big(n^\kappa e^\lambda_{d}\nabla_\kappa\big(\widetilde{E}_\lambda^{l}\Sigma_l\big)\Big)\nonumber \\
  &&+\widetilde{E}_\mu^{c}  n_\nu  K^d_{\ c}\Sigma_d-n_\mu  n_\nu a^d \Sigma_d \,,
  \end{eqnarray}
  \begin{eqnarray}
 \nabla_\mu \Omega &=& e_\mu^{\ c}D_c\Omega- n_\mu n^\kappa\nabla_\kappa \Omega  \,.
\end{eqnarray}
Replacing we obtain a general expression for the projections of the second covariant derivatives of a product of bumblebee fields
\begin{eqnarray}
\nabla_\mu\nabla_\nu\big(B^\alpha B^\beta\big) &=& \widetilde{E}_\mu^{c} e^\alpha_{a}e^\beta_{b}\widetilde{E}_\nu^{d}  \Big[D_c\Phi^{ab}_{\ \ d} -K^b_{\ c}\Theta^a_{\ d}-K^a_{\ c}\Theta^b_{\ d}-K_{cd}\Psi^{ab}\Big]\nonumber \\
&\ &- n_\mu  e^\alpha_{a}e^\beta_{b}\widetilde{E}_\nu^{d}\Big[ \widetilde{E}_\gamma^{a}\widetilde{E}_\delta^{b} e^\lambda_{d} n^\kappa\nabla_\kappa \big(e^\gamma_{c} e^\delta_{e} \widetilde{E}_\lambda^{l}\Phi^{ce}_{\ \ l}\big)-a^b\Theta^a_{\ d}-a^a \Theta^b_{\ d}-a_d \Psi^{ab}\Big]\nonumber \\
&\ &-\widetilde{E}_\mu^{c} n^\alpha  e^\beta_{b}\widetilde{E}_\nu^{d}\Big[D_c\Theta^b_{\ d}-K_{ca}\Phi^{ab}_{\ \ d}-K_{cd}\Pi^b-K^b_{\ c}\Sigma_d\Big]\nonumber \\
&\ &-\widetilde{E}_\mu^{c} e^\alpha_{a}n^\beta \widetilde{E}_\nu^{d} \Big[D_c\Theta^a_{\ d}-K_{cb}\Phi^{ab}_{\ \ d}-K_{cd}\Pi^a-K^a_{\ c}\Sigma_d\Big] \nonumber \\
&\ &-\widetilde{E}_\mu^{c} e^\alpha_{a}e^\beta_{b} n_\nu \Big[D_c\Psi^{ab}- K^d_{\ c}\Phi^{ab}_{\ \ d}-K^b_{\ c}\Pi^a-K^a_{\ c}\Pi^b\Big]\nonumber \\
&\ &+ n_\mu n^\alpha  e^\beta_{b} \widetilde{E}_\nu^{d}  \Big[ e^\lambda_{d}\widetilde{E}_\delta^{b}n^\kappa\nabla_\kappa\big(e^\delta_{e}\widetilde{E}_\lambda^{l}\Theta^e_{\ l}\big)-a_a\Phi^{ab}_{\ \ d}-a_d\Pi^b-a^b\Sigma_d\Big]\nonumber \\
&\ &+ n_\mu  e^\alpha_{a}n^\beta \widetilde{E}_\nu^{d}\Big[  e^\lambda_{d}\widetilde{E}_\gamma^{a}n^\kappa\nabla_\kappa\big(e^\gamma_{c}\widetilde{E}_\lambda^{l}\Theta^c_{\ l}\big)- a_b  \Phi^{ab}_{\ \ d}-a_d \Pi^a-a^a\Sigma_d\Big]\nonumber \\
 &\ &+n_\mu  e^\alpha_{a}e^\beta_{b} n_\nu  \Big[ \widetilde{E}_\gamma^{a}\widetilde{E}_\delta^{b}n^\kappa \nabla_\kappa\big(e^\gamma_{c}e^\delta_{d}\Psi^{cd}\big)- a^d  \Phi^{ab}_{\ \ d}-a^b \Pi^a-a^a\Pi^b\Big]\nonumber \\
 &\ &+\widetilde{E}_\mu^{c}  n^\alpha   n^\beta \widetilde{E}_\nu^{d}\Big[D_c \Sigma_d-K_{ca}\Theta^a_{\ d}-  K_{cb}\Theta^b_{\ d}-K_{mn}\Omega\Big]\nonumber \\
 &\ &+\widetilde{E}_\mu^{c}    n^\alpha e^\beta_{b}n_\nu \Big[D_c\Pi^b-K^d_{\ c}\Theta^b_{\ d}-K_{ca}\Psi^{ab}-K^b_{\ c}\Omega\Big] \nonumber \\
 &\ &+\widetilde{E}_\mu^{c}    e^\alpha_{a} n^\beta n_\nu \Big[D_c\Pi^a-K^d_{\ c}\Theta^a_{\ d}-K_{cb}\Psi^{ab}-K^a_{\ c}\Omega\Big]\nonumber \\
 &\ &- n_\mu  n^\alpha n^\beta \widetilde{E}_\nu^{d}\Big[n^\kappa e^\lambda_{d}\nabla_\kappa\big(\widetilde{E}_\lambda^{l}\Sigma_l\big)-a_a \Theta^a_{\ d}-a_b\Theta^b_{\ d}-a_d\Omega\Big]\nonumber \\
 &\ &- n_\mu n^\alpha e^\beta_{b} n_\nu \Big[\widetilde{E}_\delta^{b}n^\kappa\nabla_\kappa\big(e^\delta_{d}\Pi^d\big)-a^d\Theta^b_{\ d}-a_a\Psi^{ab}-a^b\Omega\Big]\nonumber \\
 &\ &- n_\mu    e^\alpha_{a}n^\beta n_\nu\Big[\widetilde{E}_\gamma^{a}n^\kappa \nabla_\kappa\big(e^\gamma_{c}\Pi^c\big)-a^d\Theta^a_{\ d}-a_b\Psi^{ab}-a^a\Omega\Big]\nonumber \\
 &\ &-\widetilde{E}_\mu^{c} n^\alpha n^\beta n_\nu \Big[D_c\Omega-K_{ca}\Pi^a -K_{cb}\Pi^b-K^d_{\ c}\Sigma_d\Big]  \nonumber \\
 &\ &+n_\mu n^\alpha n^\beta n_\nu  \Big[n^\kappa\nabla_\kappa \Omega -a_a\Pi^a-a_b\Pi^b-a^d\Sigma_d\Big]
\end{eqnarray}
In this way, we have successfully projected the entire tensor $\nabla_\mu\nabla_\nu(B^\alpha B^\beta)$,
which serves as the common core for constructing the bumblebee energy-momentum tensor. Each term involving second covariant derivatives can be obtained through the contractions of this tensor. This calculation considers a general bumblebee field with functional dependence on both time and hypersurface coordinates. For the specific case under study, this result will be simplified to a tangential bumblebee field and later to a time-only functional dependence. The remaining contractions and those with the Riemann tensor are left for the reader.

\end{document}